\documentclass[aps,pra,twocolumn,notitlepage,showkeys,nofootinbib,showpacs,superscriptaddress,10pt,amssymb,amsmath]{revtex4-1}
\usepackage{graphicx}
\usepackage{calc}
\usepackage[pdftex]{xcolor}
\definecolor{darkblue}{rgb}{0.0,0.0,0.6}
\definecolor{darkred}{rgb}{0.6,0.0,0}
\definecolor{lightblue}{rgb}{0.7, 0.7, 1}
\definecolor{lightred}{rgb}{1, 0.8, 0.8}
\usepackage{braket}
\usepackage[pdftex,
        colorlinks=true,
        linkcolor=darkblue,
        citecolor=darkblue,
        filecolor=black,
        urlcolor=darkblue,
        bookmarks=true,
        bookmarksopen=true,
        bookmarksopenlevel=3,
        plainpages=false,
        pdfpagelabels=true,
        verbose=true]{hyperref}

\usepackage[all]{hypcap}

\newcommand*\phantomas[3][c]{\ifmmode\makebox[\widthof{$#2$}][#1]{$#3$}\else\makebox[\widthof{#2}][#1]{#3}\fi}
\def\bea{\begin{eqnarray}}
\def\eea{\end{eqnarray}}

\newcommand{\Ho}{\hat{\mathcal{H}}}
\newcommand{\Uo}{\hat{U}}
\newcommand{\Qo}{\hat{Q}}
\newcommand{\la}{\langle}
\newcommand{\rra}{\rangle\!\rangle}
\newcommand{\lla}{\langle\!\langle}
\newcommand{\ra}{\rangle}
\DeclareMathAlphabet{\mathpzc}{OT1}{pzc}{m}{it}
\newcommand{\imu}{\mathpzc{i}}

\begin{document}

\title{Multiphoton interband excitations of quantum gases in driven optical lattices}

\author{M. Weinberg}
\author{C. \"{O}lschl\"{a}ger}
\affiliation{Institut f\"{u}r Laserphysik, Universit\"{a}t Hamburg, Luruper Chaussee 149, D-22761 Hamburg, Germany}
\author{C. Str\"{a}ter}
\author{S. Prelle}
\affiliation{Institut f\"{u}r Laserphysik, Universit\"{a}t Hamburg, Luruper Ch. 149, D-22761 Hamburg, Germany}
\author{A. Eckardt}
\author{K. Sengstock}
\author{J. Simonet}
\affiliation{Institut f\"{u}r Laserphysik, Universit\"{a}t Hamburg, Luruper Ch. 149, D-22761 Hamburg, Germany}
\affiliation{Zentrum  f\"ur Optische Quantentechnologien, Universit\"at Hamburg, Luruper Ch. 149, D-22761 Hamburg, Germany}

\begin{abstract}
    We report on the observation of multiphoton interband absorption processes for quantum gases in shaken light crystals. Periodic inertial forcing, induced by a spatial motion of the lattice potential, drives multiphoton interband excitations of up to the ninth order. The occurrence of such excitation features is systematically investigated with respect to the potential depth and the driving amplitude. \emph{Ab initio} calculations of resonance positions as well as numerical evaluation of their strengths exhibit good agreement with experimental data. In addition our findings could make it possible to reach novel phases of quantum matter by tailoring appropriate driving schemes.
\end{abstract}

\maketitle

\section{Introduction}\vspace{-3mm}
Periodic driving of quantum systems allows for the targeted engineering of exotic properties. In recent years, this approach has been very successfully utilized in various fields of physics. While time-periodic forcing of solid-state materials yields access to, e.g., topological band structures \cite{Oka2009,Lindner2011,Kitagawa2011,Wang2013} and high-$T_c$ superconductors \cite{Fausti2011,Foerst2014,Matsunaga2014,Mankowsky2014} it is also applied to trapped ions \cite{Bermudez2011}, photonic crystals \cite{Rechtsman2013} and for ultracold atomic ensembles \cite{Lignier:2007du,Kierig:2008kb}. Quantum gases in optical lattices are particularly well suited as they are almost perfectly isolated from their environment and allow for unprecedented control in a time-dependent fashion. So far, experimental studies have focused on the creation of tunable artificial gauge potentials and large effective magnetic fluxes in optical lattices \cite{Struck:2011,Aidelsburger:2011hl,Struck:2012gc,Aidelsburger:2013du,Struck:2013ar,Parker:SsK04TnP,Aidelsburger2013,Atala2014,Jotzu2014,Aidelsburger2015}, which allow for the observation of exotic phenomena such as geometrical frustration \cite{Struck:2011}, chiral Meissner currents \cite{Atala2014}, Ising magnetism \cite{Struck:2013ar,Parker:SsK04TnP,Ha2015}, and topological band structures \cite{Jotzu2014,Aidelsburger2015}. Further driving schemes have been proposed that aim for the realization of non-Abelian gauge fields \cite{Osterloh:2005bm,Hauke:2012dh}.

According to the Floquet theorem, the evolution of time-periodic systems can be described in terms of a \textit{time-periodic} unitary operator and a \textit{time-independent} effective Hamiltonian \cite{Goldman2014,Bukov2014,Goldman2015,Eckardt2015}. The underlying principle of all driving schemes is that the properties of the driven system are determined by the effective Hamiltonian, which might exhibit interesting novel features. This so-called \emph{Floquet engineering} typically assumes that excited Bloch bands can be neglected. However, quantum gases in periodically driven optical lattices exhibit close analogies with laser-irradiated solid-state materials in which nonlinear processes play a crucial role at large field strengths \cite{Nathan1985,Hemmerich1994,Sias:2008jf}. Indeed, similar to an oscillating light field, external periodic forcing of optical lattice systems is expected to induce significant multiphoton excitations between energy bands \cite{Arlinghaus:2010bz}. Thus, a deeper understanding of such excitations processes is essential for tailoring appropriate driving schemes.

Beyond its relevance to Floquet engineering, periodic forcing of optical lattices allows for the investigation of multiphoton absorption (MPA) processes in well-controlled model systems, in which the band structure and the interaction strength are fully tunable. The behavior of such non-linear excitations in interacting systems opens far-reaching lines of investigation.

In this paper we present a systematic study of multiphoton excitations of ultracold bosonic quantum gases in driven optical lattices. Periodic inertial forcing of the atomic ensemble that is induced by a rapid shaking of the rigid lattice potential results in interband excitations due to MPA of low-energy driving photons. The emergence of resonance features corresponding to MPA between different Bloch bands is investigated with respect to the lattice depth and the driving field intensity in a one dimensional optical lattice. We extend our experimental studies to a two-dimensional triangular lattice in the regime of negative effective tunneling. The positions of the observed resonances as well as their relative strengths are in good agreement with theoretical simulations.

\begin{figure}[t]\hypertarget{fig01ht}{}
\centering
\includegraphics{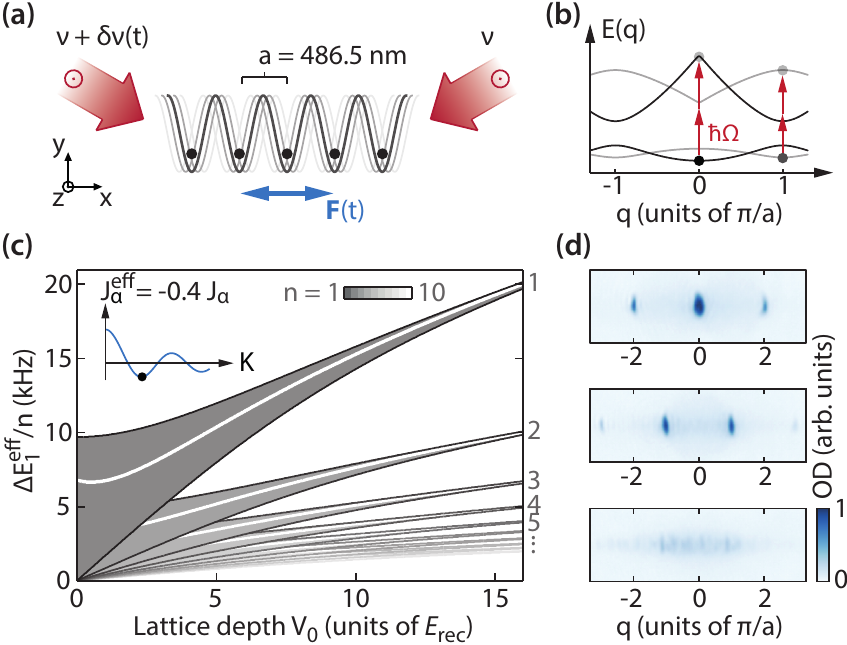}
\caption{(Color online) (a) Setup of the driven 1D lattice with both beams being linearly polarized along the $z$-axis. (b) Illustration of two-photon transitions into the first excited band at the minima at $q=0$ for $J^{\mathrm{eff}}_{0}>0$ (black) and $q=\pi/a$ for $J^{\mathrm{eff}}_{0}<0$ (gray). (c) Multiphoton transition energies (white lines) to the first excited band according to Eq.\,(\ref{eq3}) for $K=3.82$ (see inset). Gray shaded areas depict the maximum possible range of MPA. (d) Typical time-of-flight images of the driven 1D lattice for positive tunneling (top panel), negative tunneling (middle panel), and a heated, incoherent sample (bottom panel).}
\label{fig1}
\end{figure}

\vspace{-2mm}\section{Experimental setup}\vspace{-3mm}
Here, multiphoton interband excitations are investigated in an ensemble of ultracold bosonic $^{87}\mathrm{Rb}$ atoms that is confined in a red-detuned one-dimensional optical lattice. As depicted in Fig.\,\hyperlink{fig01ht}{\ref{fig1}(a)}, the lattice consists of a pair of running-wave laser beams of wavelength $\lambda_L=830\,\mathrm{nm}$ that are arranged at an angle of $117.1^{\circ}$ with respect to each other in the \emph{xy}-plane. The resulting standing light wave $V_{\mathrm{lat}}(x)=V_0\cos(2\pi x/a)/2$ has a lattice spacing of $a=486.5\,\mathrm{nm}$. Its potential depth $V_0$ is conveniently expressed in units of the recoil energy $E_{\mathrm{rec}}=\hbar^2k_L^2/(2M)$, denoting the kinetic energy that is transferred to an atom of mass $M$ by absorbing a lattice photon with wavenumber $k_L=2\pi/\lambda_L$, where $\hbar$ is the reduced Planck constant. In the following, the band structure of the lattice is denoted by $\varepsilon_\alpha(q)$, with band index $\alpha=0,1,2,\ldots$ and quasimomentum wave number $q\in(-\pi/a,\pi/a]$ in the $x$-direction. The lowest bands are dominated by tunneling between neighboring lattice minima with tunneling parameters $J_0>0$ and $J_1<0$, respectively, so that $\varepsilon_0(q)\approx-2 J_0\cos(aq)$ and $\varepsilon_1(q)\approx \bar{\varepsilon}_1 - 2J_1\cos(aq)$, with $\bar{\varepsilon}_1$ being the band-center energy.

\vspace{-2mm}\section{Periodic driving}\vspace{-3mm}
Periodic driving of the system is induced by a sinusoidal frequency modulation $\delta\nu(t)=\nu_0\sin(\Omega t)$ of one of the two laser beams. This modulation gives rise to a periodic motion of the potential along the lattice axis. In the co-moving frame, the atoms experience an inertial force $\mathbf{F}(t)=F_0\cos(\Omega t)\hspace{0.5mm}\mathbf{\hat{e}}_x$ with an amplitude of $F_0=M\Omega\nu_0a$. Apart from trap and interactions, the neutral particles are described by the Hamiltonian
\begin{align}\label{eq1nn}
\hat{\mathcal{H}}(t) = \frac{\mathbf{\hat{p}}^2}{2M}+V_\text{lat}(x)+xF_0\cos(\Omega t).
\end{align}
The driving term in Eq.\,(\ref{eq1nn}) breaks the translational symmetry of the lattice which can be restored by the gauge transformation described in Appendix \ref{AppendixA} \cite{Eckardt:2010fo,Arimondo:2012kf}. In the new reference frame of the lattice the resulting Hamiltonian can be written as
\begin{align}\label{eq1}
\hat{\mathcal{H}}'(t) = \frac{\mathbf{[\hat{p}-A(t)]}^2}{2M}+V_\text{lat}(x),
\end{align}
where the effect of the periodic driving is incorporated into a time-dependent vector potential $\mathbf{A}(t)$ given via the relation $\mathbf{F}(t)=-\partial\mathbf{A}(t)/\partial t$. In the basis of static Bloch states $|\alpha\phantom{'} q\rangle$, this yields a tight-binding Hamiltonian
\begin{align}\label{eq11}
\begin{split}
\hat{\mathcal{H}}'_q(t)=\sum_\alpha\Big[\varepsilon_\alpha\big(q-A(t)/\hbar\big)&|\alpha\phantom{'} q\rangle\langle\alpha\phantom{'} q|\\
+aF_0\cos(\Omega t)\sum_{\alpha'}\eta_{\alpha,\alpha'}&|\alpha' q\rangle\langle\alpha\phantom{'} q|\Big],
\end{split}
\end{align}
Here, and $A(t)\equiv A_x(t)$ and $\eta_{\alpha,\alpha'}$ are dimensionless dipole matrix elements coupling bands $\alpha$ and $\alpha'$ defined in Appendix \ref{AppendixA}.

For driving frequencies that are large compared to the widths of the bands, the impact of periodic forcing can be understood as a combination of two effects as detailed in Appendix \ref{AppendixB}. The first effect is a modification of each single band, described by the time-averaged single-particle dispersion relation
\begin{align}\label{eq2}
\varepsilon^\mathrm{eff}_\alpha(q)&=\frac{1}{T}\int_0^T\!dt\,\varepsilon_\alpha\big(q-A(t)/\hbar\big)\\
&= \bar{\varepsilon}_\alpha -2J_\alpha{\mathcal{J}_0}(K)\cos(aq).
\end{align}
This yields the effective modification of nearest-neighbor tunneling matrix elements $J_\alpha^\mathrm{eff}=J_\alpha {\mathcal{J}}_0(K)$, where $\mathcal{J}_0$ is the zeroth order Bessel function of the first kind and $K=aF_0/(\hbar\Omega)$ denotes a dimensionless driving amplitude. The second effect, resulting from the second term in Eq.\,(\ref{eq11}), can be understood as the resonant coupling of the so-modified bands.

Experimentally, the driving amplitude $K$ is linearly increased to its final value within $50\,\mathrm{ms}$ after the atomic ensemble is prepared in the lowest-energy band of the optical lattice. Driving is maintained at the final amplitude for another $20\,\mathrm{ms}$. Subsequently, all trapping potentials are rapidly switched off, and atoms fall freely under the influence of gravity for $40\,\mathrm{ms}$ time-of-flight before a resonant absorption image is taken.

For the lattice depths used throughout the presented experiments the atomic ensemble remains in the weakly interacting superfluid regime. Thus, the inversion of the effective band structure due to the sign change of $J_\alpha^\mathrm{eff}$ for sufficiently large forcing amplitudes [see Fig.\,\hyperlink{fig01ht}{\ref{fig1}(b)}] can be identified by the position of the coherent quasimomentum peaks in time-of-flight absorption images as shown in the upper two panels in Fig.\,\hyperlink{fig01ht}{\ref{fig1}(d)}.

\vspace{-2mm}\section{Interband multiphoton transitions}\vspace{-3mm}
As explained in the previous section, beyond the tunneling renormalization, the periodic forcing of frequency $\Omega$ also induces finite matrix elements for coherent interband coupling processes that conserve quasimomentum but allow for energy changes of integer multiples of the ``photon'' energy $\hbar\Omega$ \cite{weinbergSM}. Hence, an $n^{\mathrm{th}}$-order multiphoton transition is expected to occur when the resonance condition
\begin{align}\label{eq3}
    n\times\hbar\Omega =
    \begin{cases}
        \hspace{1mm}\Delta E^{\mathrm{eff}}_{\alpha}(q=0)&\text{for}\hspace{3mm}J^{\mathrm{eff}}_0>0 \\
        \hspace{1mm}\Delta E^{\mathrm{eff}}_{\alpha}(q=\pm\pi/a)&\text{for}\hspace{3mm}J^{\mathrm{eff}}_0<0
    \end{cases}
\end{align}
with $\Delta E^{\mathrm{eff}}_{\alpha}(q) =  \varepsilon^\mathrm{eff}_{\alpha}(q)- \varepsilon^\mathrm{eff}_0(q)$ is fulfilled. Condition (\ref{eq3}) is plotted in Fig.\,\hyperlink{fig01ht}{\ref{fig1}(c)} for $\alpha=1$ and $n=1,\ldots ,10$ (white lines) as a function of the lattice depth for a driving amplitude of $K=3.82$ where $J^{\mathrm{eff}}_{\alpha}=-0.4J_{\alpha}$ (see inset).

These interband transitions, induced by the periodic driving, significantly reduce the maximum optical density extracted from the time-of-flight images as can be observed in the bottom panel of Fig.\,\hyperlink{fig01ht}{\ref{fig1}(d)} \cite{weinbergSM}. Such a decrease in the visibility in the absorption images can be ascribed to two distinct processes. First, further MPA might populate higher lying bands that are no longer trapped in the optical lattice and, thus, result in a decrease of the optical density due to atomic losses. Second, interacting Bose-Einstein condensates in excited bands rapidly decay due to scattering processes, thereby reducing the degree of coherence in the system \cite{Martikainen2011,Paul2013}.

\begin{figure}[t]\hypertarget{fig02ht}{}
\centering
\includegraphics{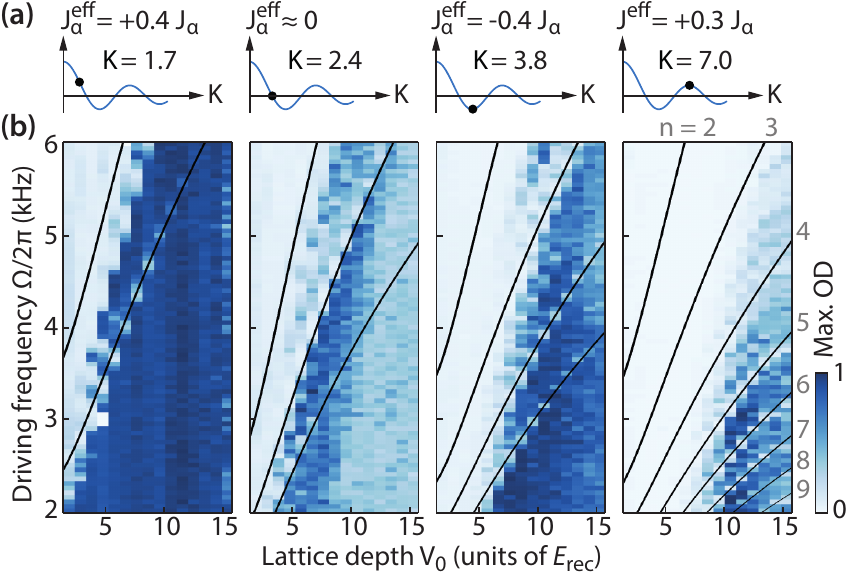}
\caption{(Color online) Systematic investigation of multiphoton spectra in the driven 1D lattice. (a) The effective tunneling matrix element $J^{\mathrm{eff}}$ acquires different values at the four measured driving amplitudes $K$, scaling with the Bessel function $\mathcal{J}_0(K)$. (b) Excitation spectra for the four driving amplitudes depicted in (a) with the maximum optical density encoded in brightness. Solid black lines, numbered on the right hand side, indicate the calculated positions of MPA to the first excited energy band according to Eq.\,(\ref{eq3}).}
\label{fig2}
\end{figure}

A spectroscopic study of these multiphoton transitions is shown in Fig.\,\ref{fig2} where excitation spectra are obtained for various lattice depths at four different driving amplitudes. The resulting effective tunneling parameters $J^{\mathrm{eff}}_{\alpha}$ [see Fig.\,\hyperlink{fig02ht}{\ref{fig2}(a)}] yield different transition energies according to Eq.\,(\ref{eq3}) that are plotted as solid black lines together with the excitation spectra in Fig.\,\hyperlink{fig02ht}{\ref{fig2}(b)}. These \emph{ab initio} calculations of transition energies exhibit excellent agreement with the experimental data. With increasing driving amplitude, higher-order excitations appear in the spectrum and the width of the resonances increases. While no excitations are present in the system above the third-order resonance for $K=1.7$, up to the ninth order multiphoton transition can be identified at a driving amplitude of $K=7.0$, where the lower orders of MPA transitions already overlap significantly. One can also observe that resonance features become weaker for increasing lattice depth; this is a consequence of the fact that the coupling matrix elements $\eta_{\alpha,\alpha'}$ become smaller for deeper lattices. Each of the four individual data sets has been normalized, yielding comparable results for all investigated driving amplitudes. Even for $K=2.4$, where the effective tunneling amplitude vanishes, the signal-to-noise ratio is sufficient to clearly identify resonance features despite a significantly reduced level of coherence.

\begin{figure}[t]\hypertarget{fig03ht}{}
\centering
\includegraphics{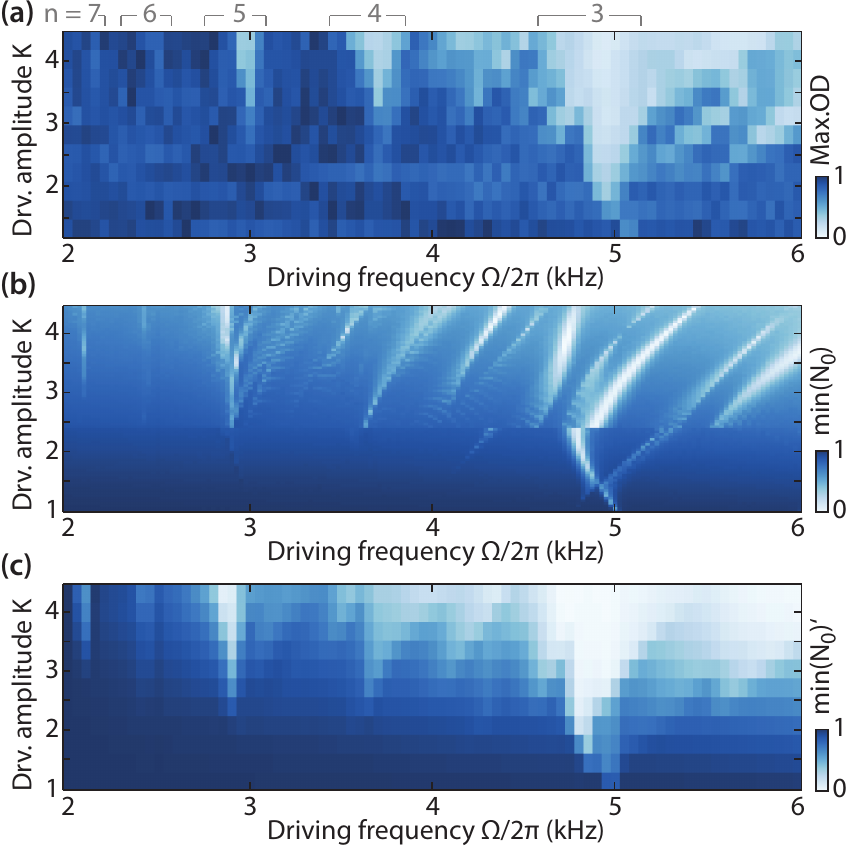}
\caption{(Color online) Emergence of multiphoton interband transitions for increasing driving amplitude. (a) Multiphoton excitation spectrum obtained at a fixed 1D lattice depth of $9.5\,E_{\mathrm{rec}}$ for increasing driving amplitudes. The positions and maximum possible widths of the depicted $n^{\mathrm{th}}$-order transitions are indicated above the spectrum. Data at $K\approx4.0$ (second row from top) have not been measured and are interpolated. (b) Numerical simulation of the observed multiphoton interband excitation spectra. Plotted is the minimum occupation $N_0$ of the lowest band observed during 20\,ms of driving at the final value of $K$. Avoided crossings and excitations to higher energy bands are clearly visible. In (c) the resolution of the simulation data shown in (b) has been reduced to match the resolution of the experimental data. To take into account the linear ramping to the final driving amplitude $K$, excitations present for smaller values of $K$ are kept in the spectra for larger $K$. The resulting spectrum clearly matches the experimental data depicted in (a).}
\label{fig3}
\end{figure}

In order to gain a deeper understanding of the emergence of interband MPA processes, the observed excitations are further explored with respect to the driving amplitude in Fig.\,\ref{fig3}. The excitation spectra depicted in Fig.\,\hyperlink{fig03ht}{\ref{fig3}(a)} are obtained at a fixed one-dimensional (1D) lattice depth of $V_0=9.5\,E_{\mathrm{rec}}$ while the driving amplitude $K$ is gradually increased. Here, we observe that for increasing final coupling strengths, the widths of the observed resonance features increase. Small additional features of the excitation spectrum in the region of large driving amplitudes $K$ and frequency $\Omega$ that cannot be explained by the expected resonance positions indicate the occurrence of multiphoton excitations to even higher-energy bands.

A numerical simulation of the interband excitation spectra, described in more detail in Appendix \ref{AppendixD}, is depicted in Fig.\,\hyperlink{fig03ht}{\ref{fig3}(b)}. Starting from an ensemble of states with quasimomenta distributed sharply around the minimum of the effective dispersion relation at the respective maximum value of $K$, the time evolution under the time dependent Hamiltonian given in Eq.(\ref{eq1}) has been integrated over $20\,\mathrm{ms}$. Excitations to higher bands are quantified by the minimum fraction of occupation $N_0$ of the lowest band during the time evolution. The obtained spectra exhibit good agreement with the experimental data in Fig.\,\hyperlink{fig03ht}{\ref{fig3}(a)}, reproducing the observed multiphoton transitions to the first excited band with $n=3$ to $7$ as well as additional resonance features that are associated with multiphoton excitations to higher lying bands at larger driving frequencies. Here, the finite width of the momentum distribution (resulting from thermal fluctuations, interaction-induced quantum fluctuations, and the trap potential) plays a central role, since it is required to explain the observed resonances of even photon number between the two lowest bands (see Appendix \ref{AppendixC}). Avoided-crossing-type structures, indicating the resonant hybridization of the first excited band with higher-lying ones, can be identified in the simulated resonance features at the $n=3,\,4$, and $5$ transitions.

To facilitate the comparison of the rich spectrum obtained from numerical simulations and the experimental data, we have reduced the resolution of the numerical data to the experimental resolution in Fig.\,\hyperlink{fig03ht}{\ref{fig3}(c)}. To emulate the linear ramping of the driving amplitude to its respective final value, each row $n$ of this plot includes the combined product of all rows corresponding to smaller values of $K$ according to $\min[N_0(K_n)]' = \prod_{i=1}^n\min[N_0(K_i)]$. With this, an excitation present during any point of the ramp remains in the system at any larger value of the driving amplitude $K$. The resulting spectrum clearly reproduces the features of the experimental data shown in Fig.\,\hyperlink{fig03ht}{\ref{fig3}(a)} such that the observed broadening of the multiphoton resonances can be ascribed to the linear ramping procedure of the driving amplitude $K$.

\vspace{-2mm}\section{Interband excitations in the triangular lattice}
\begin{figure}[t]\hypertarget{fig04ht}{}
\centering
\includegraphics{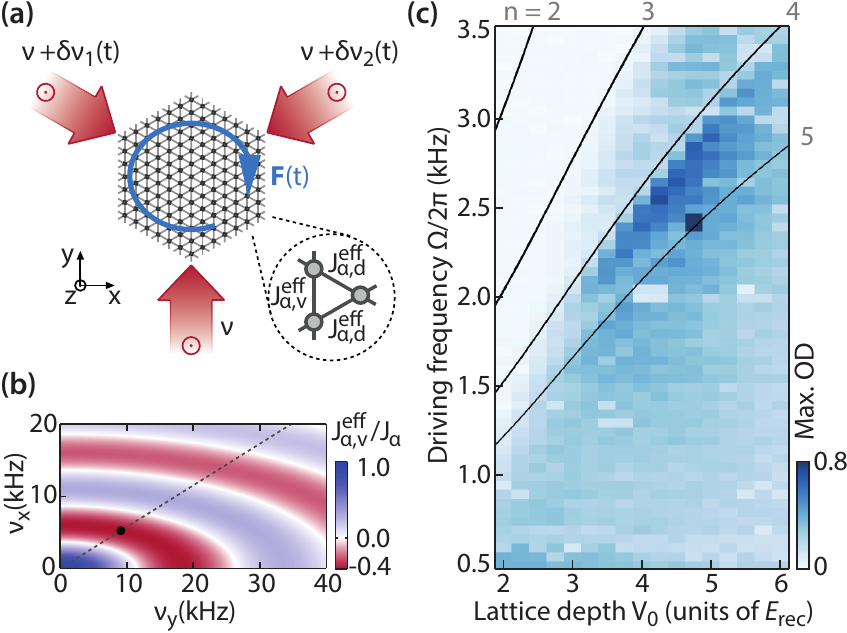}
\caption{(Color online) Driving in the triangular lattice. (a) Illustration of the experimental setup, lattice structure and elliptical forcing. (b) Tunneling renormalization of the diagonal lattice bonds, which depend on the horizontal and vertical frequency modulation amplitudes $\nu_{x/y}$. The isotropic renormalization condition of $\nu_{y} = \sqrt{3}\nu_{x}$ is plotted as a dashed line \cite{weinbergSM}. (c) Excitation spectrum for a fixed isotropic driving amplitude of $K=3.82$, indicated by the solid dot in (b). Expected positions and widths for multiphoton transitions are plotted as solid black lines similar to Fig.\,\ref{fig2}. MPA resonances of up to the fifth order can be clearly identified.}
\label{fig4}
\end{figure}

\vspace{-3mm}
For the investigation of interband MPA processes in more complex lattice structures we extend our studies to a driven two-dimensional triangular lattice. As depicted in Fig.\,\hyperlink{fig04ht}{\ref{fig4}(a)} the lattice is composed of three running wave laser beams of equal intensity intersecting in the $xy$-plane with linear out-of-plane polarizations \cite{weinbergSM}.  Here, inertial forcing is induced by a sinusoidal frequency modulation of two of the three laser beams, resulting in a periodic elliptical forcing of the rigid lattice potential. This allows adjusting the amplitude and sign of two effective tunneling directions, denoted $J^{\mathrm{eff}}_{\alpha,\mathrm{v}}$ and $J^{\mathrm{eff}}_{\alpha,\mathrm{d}}$ in Fig.\,\hyperlink{fig04ht}{\ref{fig4}(a)}, independently.  The tunneling renormalization of $J^{\mathrm{eff}}_{\alpha,\mathrm{d}}$ is shown in Fig.\,\hyperlink{fig04ht}{\ref{fig4}(b)} to depend on the horizontal and vertical frequency modulation components $\nu_{x}$ and $\nu_{y}$, which determine the ratio of the major and minor axes of the elliptical forcing $\mathbf{F}(t)$ \cite{weinbergSM}. For the investigation of MPA in the driven triangular lattice, we focus on an isotropic forcing parameter of $K=3.82$ along all lattice bonds. This corresponds to a negative effective tunneling of maximal amplitude ($J^{\mathrm{eff}}_{\alpha,\mathrm{v}}=J^{\mathrm{eff}}_{\alpha,\mathrm{d}}=-0.4J_{\alpha}$), which is central for the study of frustrated magnetism \cite{Struck:2011}.

MPA resonances of up to the fourth order can be clearly identified in the excitation spectrum shown in Fig.\,\hyperlink{fig04ht}{\ref{fig4}(c)} with excellent agreement with the calculated transition energies to the first excited Bloch band (solid black lines). In addition, faint remnants of a fifth order transition are visible below lattice depths of $4.5\,E_{\mathrm{rec}}$. The maximum coherence of the atomic ensemble is reached only at a narrow parameter region between the fourth and the fifth order transitions while the maximum optical density remains small for smaller driving frequencies where heating due to the resonant creation of collective intraband excitations might occur  \cite{Eckardt2015}. Such a significant limitation of the accessible parameter space for coherent manipulation of atomic ensembles is a crucial aspect for the experimental realization of periodic driving schemes in two- or higher-dimensional lattice systems that rely on the applicability of time-averaged effective models. In addition, avoiding possible MPA processes is even more demanding for driving schemes employing more than a single driving frequency \cite{Verdeny2015}.

\vspace{-2mm}\section{Conclusion}\vspace{-3mm}
To conclude, multiphoton interband excitations have been investigated systematically with ultracold quantum gases in optical lattices. Thereby, multiphoton transitions to the first excited energy band of up to the ninth order could be observed in a driven one-dimensional lattice as well as in a two-dimensional triangular lattice. The resonance positions are found to be in excellent agreement with \emph{ab initio} calculations of the time-averaged effective single-particle band structure. Also, the strength of the resonances and their dependence on the driving amplitude show good agreement with numerical simulations.

Our findings provide essential insights concerning the applicability of strong driving schemes for the experimental realization of exotic quantum phases in the rapidly growing field of Floquet engineering. Moreover, a comprehensive understanding of driven mesoscopic matter waves is a crucial prerequisite for the coherent control and addressability of intriguing quantum states. For instance, in analogy to coherent light-matter interactions, external periodic driving with precisely adjusted pulse shapes could allow for the generation of cat-like states between different Bloch bands in optical lattices \cite{Arlinghaus:2011cy}. In addition, the particularly strong inertial forcing enables the emulation of extremely strong field conditions in condensed matter systems that are hardly accessible with real solids \cite{Arlinghaus:2010bz}. A further strength of quantum gases relies on the precise control over the interactions in the system. It is essential for the investigation of the complex interplay between periodic driving and interactions, which has very recently triggered several theoretical studies \cite{Choudhury2014,DAlessio2014,Choudhury2015,Straeter2015,Bilitewski2015}.

\vspace{-2mm}\section*{Acknowledgments}\vspace{-3mm}
We thank M.~Holthaus for fruitful discussions. This work was funded by the SFB925 of the Deutsche Forschungsgemeinschaft and the European Community’s Seventh Framework Programme (FP7/2007-2013) under grant agreement No.\,323714 (EQuaM).

\begin{appendix}

\section{The driven lattice system}
\label{AppendixA}

In the following we provide a theory of multiphoton interband transitions in the driven 1D lattice and will identify the basic processes that lead to multi-``photon'' interband transitions in a driven optical lattice and their rates.

In the lattice frame of reference a sinusoidal shaking of the lattice gives rise to a time-periodic homogeneous inertial force, resembling an ac voltage. Thus, a particle of mass $M$ in a shaking cosine lattice is described by Eq.\,(\ref{eq1nn}) of the main text:
\begin{equation}\label{eq:s12}
\Ho(t) = -\frac{\hbar^2}{2M}\partial_x^2 - \frac{V_0}{2}\cos(2\pi x/a) + x F_0 \cos(\Omega t),
\end{equation}
with $V_0$ being the lattice depth, $a$ being the lattice constant, and $F_0$ being the amplitude of the force.

It is convenient to describe the driven lattice in terms of the maximally localized Wannier states $|\ell \alpha\ra$ of the undriven Hamiltonian ($F_0=0$). The integer $\ell$ labels the lattice minima $x_\ell=\ell a$ and the index $\alpha=0,1,2,\ldots$ denotes the Blochs bands spanned by the corresponding Wannier states, with the energy increasing with $\alpha$. The Wannier wave functions $w_\alpha(x-\ell a)=\la x|\ell \alpha\ra$ are real, exponentially localized on a length increasing with $\alpha$, symmetric (antisymmetric) for even (odd) bands, $w_\alpha(-x)=(-)^\alpha w_\alpha(x)$, and shall be normalized, $\int\, d x \,|w_\alpha(x)|^2=1$.

In Wannier representation, non-interacting particles in the driven lattice are described by the tight-binding Hamiltonian
\begin{eqnarray}\label{eq:Htb}
\Ho(t)&=& \sum_{\ell\alpha}\bigg[ \varepsilon_\alpha  |\alpha\ell\ra\la\alpha\ell| 
- J_\alpha \Big(  |\alpha(\ell+2)\ra \la\alpha \ell| +\text{h.c.} \Big)
\nonumber\\&&
		+\,\tilde{K} \cos(\Omega t) \Big( \ell |\alpha\ell\ra\la\alpha\ell| 
			+ \sum_{\alpha'}\eta_{\alpha'\alpha} |\alpha'\ell\ra\la\alpha\ell| 				
								\Big)\bigg].
\nonumber\\
\end{eqnarray}
The Hamiltonian is characterized by the amplitude of the potential modulations $\tilde{K}=F_0a$, the band-center energies

\begin{align}
\begin{split}
\bar{\varepsilon}_\alpha=\int\,dx\, w_\alpha(x&)\Big[-\frac{\hbar^2}{2M}\partial_x^2 \\&- \frac{V_0}{2}\cos(2\pi x/a)\Big]w_\alpha(x),
\end{split}
\end{align}
the nearest-neighbor tunneling parameters
\begin{align}
\begin{split}
J_\alpha= -\hspace{-1mm}\int\!dx\, w_\alpha(x-a&)\Big[-\frac{\hbar^2}{2M}\partial_x^2 \\&- \frac{V_0}{2}\cos(2\pi x/a)\Big]w_\alpha(x),
\end{split}
\end{align}
and the dimensionless dipole matrix elements
\begin{equation}
\eta_{\alpha'\alpha}=\frac{1}{a}\int\!dx\, w_{\alpha'}(x)x w_\alpha(x)
\end{equation}
describing interband coupling. As a consequence of the parity of the Wannier functions, $\eta_{\alpha'\alpha}$ is non zero only if $(\alpha'-\alpha)$ is odd. Moreover the sign of the tunneling parameter $J_\alpha$ alternates with the band index $J_\alpha/|J_\alpha|=(-)^\alpha$. It is an approximation to neglect tunneling beyond nearest-neighbor sites and not to take into account time-periodic band-coupling terms between Wannier states on different lattice sites. However, for the two low-lying bands those terms are rather small. The neglected terms might still be relevant for the description of higher-lying bands though. Including the neglected terms in the analysis presented below would be straightforward, but is not done here for the sake of a simple presentation, capturing the basic picture. However, these terms are included in our numerical simulation of interband transitions described in Appendix \ref{AppendixD}.

The driving term breaks the translational symmetry of the lattice. However, the symmetry can be restored by performing a gauge transformation
\begin{equation}\label{eq:Hprime}
\Ho'(t)=\Uo^\dag(t)\Ho(t)\Uo(t)-\imu\hbar\Uo^\dag(t){\dot\Uo}(t),
\end{equation}
with the unitary operator given by
\begin{equation}
\Uo(t) = \exp\Big(\imu\sum_{\ell\alpha}\chi_\ell(t)|\alpha\ell\ra\la\alpha\ell|\Big) 
\end{equation}
and
\begin{equation}
\chi_\ell(t)=-\frac{\tilde{K}\ell}{\hbar}\int_0^t\!dt' \cos(\Omega t') =  -\ell K \sin(\Omega t),
\end{equation}
where $K=\tilde{K}/(\hbar\Omega)$. This unitary transformation integrates out the oscillatory shift in quasimomentum by
\begin{equation}
\Delta q(t)= -\frac{K}{a}\sin(\Omega t)
\end{equation}
induced by the periodic force. The new Hamiltonian reads
\begin{eqnarray}
\Ho'(t) &=& \sum_{\ell\alpha} \bigg[ \bar{\varepsilon}_\alpha |\alpha\ell\ra\la\alpha\ell| 
		-J_\alpha \big( e^{\imu\theta(t)} |\alpha(\ell+1)\ra\la\alpha\ell| + \text{H.c.}\big)
\nonumber\\&&
	+\,\tilde{K}\cos(\Omega t) \sum_{\alpha'}\eta_{\alpha'\alpha} |\alpha'\ell\ra\la\alpha\ell| \bigg]
\end{eqnarray}
with time-periodic Peierls phase $\theta(t)=\chi_{\ell}(t)-\chi_{\ell+1}(t)=-a\Delta q(t)=K\sin(\Omega t)$.

Let us, for simplicity, assume a system of $N$ lattice sites under periodic boundary conditions and express the Hamiltonian in terms of Bloch states $|\alpha q\ra$, with the quasimomentum quantum number $q$ taking $N$ discrete values $k=2\pi\mu/(Na)$, with integer $\mu$ in the interval $(-\frac{\pi}{a},\frac{\pi}{a}]$. Using $\la\alpha' \ell|\alpha q\ra=\delta_{\alpha'\alpha}N^{-1/2}\exp(\imu\ell a q)$, one finds
\begin{equation}\label{eq:Hprimek}
\Ho'(t)= \sum_{q} \Ho'_q(t),
\end{equation}
with
\begin{eqnarray}
\Ho'_q(t) &=&
\sum_\alpha\bigg[\varepsilon_\alpha\big(q-A(t)/\hbar\big)|\alpha q\ra\la\alpha q|
\nonumber\\&&
		+\,\tilde{K}\cos(\Omega t) \sum_{\alpha'}\eta_{\alpha'\alpha} |\alpha'q\ra\la\alpha q|\bigg].
\end{eqnarray}
corresponding to Eq.\,(\ref{eq11}) in the main text. Here,
\begin{equation}
A(t)=-\hbar\Delta q(t)
\end{equation}
plays the role of a vector potential and
\begin{equation}
\varepsilon_\alpha(q) = \bar{\varepsilon}_\alpha - 2 J_\alpha \cos(qa)
\end{equation}
denotes the single-particle dispersion relation for band $\alpha$.

As a consequence of the discrete translational invariance, the Hamiltonian $\Ho'(t)$ conserves quasimomentum $q$. That means that interband transitions will occur between Bloch states $|\alpha q\ra$ and $|\alpha' q\ra$ of the same quasimomentum $q$. This reduces the problem to independent subspaces that are characterized by $q$ and spanned by states labeled by the band index $\alpha$.

\section{Floquet picture}
\label{AppendixB}
Let us investigate the problem within the extended Floquet Hilbert space. The time-dependent Schr\"odinger equation
\begin{equation}
\imu\hbar d_t|\psi(t)\ra =\Ho'(t)|\psi(t)\ra
\end{equation}
possesses a set of generalized stationary states of the form
\begin{equation}
|\psi_\nu(t)\ra = |u_\nu(t)\ra e^{-\frac{\imu}{\hbar} t \varepsilon_\nu }
\end{equation}
called Floquet states. They are characterized by the time-periodic Floquet mode $|u_\nu(t)\ra=|u_\nu(t+T)\ra$ and by the quasienergy $\varepsilon_\nu$, where the driving period $T$ is defined by $T=2\pi/\Omega$. The Floquet states, which are labeled by some quantum number $\nu$, form a complete basis of the state space at any time $t$. Therefore, we can expand the time evolution of a state $|\psi(t)\ra$ in terms of the Floquet states like
\begin{equation}
|\psi(t)\ra = \sum_\nu c_\nu|\psi_\nu(t)\ra
=  \sum_\nu c_\nu |u_\nu(t)\ra e^{-\frac{\imu}{\hbar} t \varepsilon_\nu }
\end{equation}
with time-independent coefficients $c_n=\la\psi_\nu(t_0)|\psi(t_0)\ra$.

Unlike the Floquet states, the quasienergies and the Floquet modes are not defined uniquely. For each Floquet state a
whole family of Floquet modes and quasienergies,
\begin{equation}
\varepsilon_{\nu m}=\varepsilon_\nu+m\hbar\Omega,
\quad
|u_{\nu m}(t)\ra = |u_{\nu}(t)\ra e^{\imu m\Omega t},
\end{equation}
labeled by the integer $m$, can be defined such that
\begin{equation}
|\psi_\nu(t)\ra = |u_{\nu m}(t)\ra e^{-\frac{\imu}{\hbar} t \varepsilon_{\nu m} }
\end{equation}
for all $m$. However, Floquet modes and quasienergies of different $m$ still constitute independent solutions of the eigenvalue problem
\begin{equation}
\bar{Q} |u_{\nu m}\rra = \varepsilon_{\nu m}|u_{\nu m}\rra
\end{equation}
of the quasienergy operator
\begin{equation}
\Qo(t)=\Ho'(t)-\imu\hbar d_t .
\end{equation}
This eigenvalue problem is defined in the extended Floquet Hilbert space, being the product space of the state space with the space of time-periodic functions. In this space the scalar product is given by
\begin{equation}
\lla u|v\rra = \frac{1}{T}\int_0^T\! dt\, \la u(t)|v(t)\ra.
\end{equation}
When considering a periodically time dependent state $|u(t)\ra=|u(t+T)\ra$ as an element of the extended Hilbert space, we write it as a double-ket $|u\rra$. Likewise, an operator is marked by an overbar, like $\bar{Q}$, if it is considered to act in the extended Hilbert space.

For the driven lattice a useful set of basis states spanning the extended Hilbert space is given by
\begin{equation}
|\alpha q m \rra: \quad |\alpha q m(t)\ra = |\alpha q\ra e^{\imu m\Omega t}
\end{equation}
and labeled by the band index $\alpha$, the quasimomentum $q$, and the Fourier index $m$. These states are the Floquet modes of the undriven problem with $\tilde{K}=0$.  With respect to these basis states the quasienergy operator possesses the matrix elements
\begin{equation}
\lla \alpha' q' m'| \bar{Q}|\alpha q m\rra
= \la \alpha' q'|\big(\Ho'_{m'-m} +\delta_{m'm}\hbar\Omega\big)|\alpha q\ra,
\end{equation}
where
\begin{equation}
\Ho'_m = \frac{1}{T} \int_0^T\! dt\, e^{- \imu m\Omega t}\Ho'(t)
\end{equation}
denotes the Fourier transform of the Hamiltonian such that $\Ho'(t)=\sum_m e^{\imu m\Omega t}\Ho_m'$. The quasienergy operator assumes a block structure with respect to the index $m$, which plays the role of a relative photon number. The diagonal blocks describe subspaces of different ``photon'' number and are shifted relative to each other in energy by integer multiples of the photon energy $\hbar\Omega$. The diagonal blocks are coupled by off-diagonal blocks characterized by $\Ho_{m\ne0}'$ describing $m$-photon processes.

Let us evaluate the matrix elements explicitly. One finds
\begin{equation}
\Ho'_m = \sum_{q\alpha} \varepsilon_{\alpha m}(q) |\alpha q\ra\la\alpha q|
			+ \frac{1}{2}\tilde{K}\delta_{|m|,1}\eta_{\alpha'\alpha}|\alpha' q\ra\la\alpha q|,
\end{equation}
where
\begin{eqnarray}
\varepsilon_{\alpha m}(q) &=& \frac{1}{T}\int_0^T\! dt \, \varepsilon_\alpha\big(q-A(t)/\hbar\big),
\nonumber\\&=& \bar{\varepsilon}_\alpha\delta_{m,0}
					- J_{\alpha}\mathcal{J}_m(K)\Big[e^{-\imu aq}+(-)^m e^{\imu aq}\Big],
\nonumber\\
\end{eqnarray}
with $\mathcal{J}_m$ denoting a Bessel function of order $m$. The diagonal blocks are given by
\begin{equation}
\lla \alpha' q' m| \bar{Q}|\alpha q m\rra
	= \delta_{q'q}\delta_{\alpha'\alpha} \big[\varepsilon^\text{eff}_\alpha(q) + m\hbar\Omega\big],
\end{equation}
where we have introduced the effective dispersion relation
\begin{equation}
\varepsilon^\text{eff}_\alpha(q) = \varepsilon_{\alpha 0}(q)
	=\bar{\varepsilon}_\alpha - 2 J_\alpha \mathcal{J}_0(K)\cos(aq).
\end{equation}
The off-diagonal blocks read
\begin{eqnarray}
\lla \alpha' q' m'| \bar{Q}|\alpha q m\rra
&=& \delta_{q'q} \Big[\delta_{\alpha'\alpha}\varepsilon_{\alpha (m'-m)}(q)
\nonumber\\&&
	+\, \frac{1}{2}\tilde{K}\delta_{|m'-m|,1}\eta_{\alpha'\alpha}\Big].
\end{eqnarray}
The first term describes multi-photon processes with arbitrary $|m'-m|$ that do not change the band index $\alpha$. The second term describes single-photon transitions with $|m'-m|=1$ between two bands $\alpha$ and $\alpha'$ with odd $(\alpha'-\alpha)$. Thus, the quasienergy operator does not directly possess matrix elements that describe multiphoton interband transitions. Such processes emerge, however, from-higher order virtual processes in the extended Floquet Hilbert space, as we will discuss in Appendix \ref{AppendixC}.

\section{Multiphoton interband coupling}\label{AppendixC}
\begin{figure}[t]\hypertarget{fig05ht}{}
\includegraphics[width=0.8\linewidth]{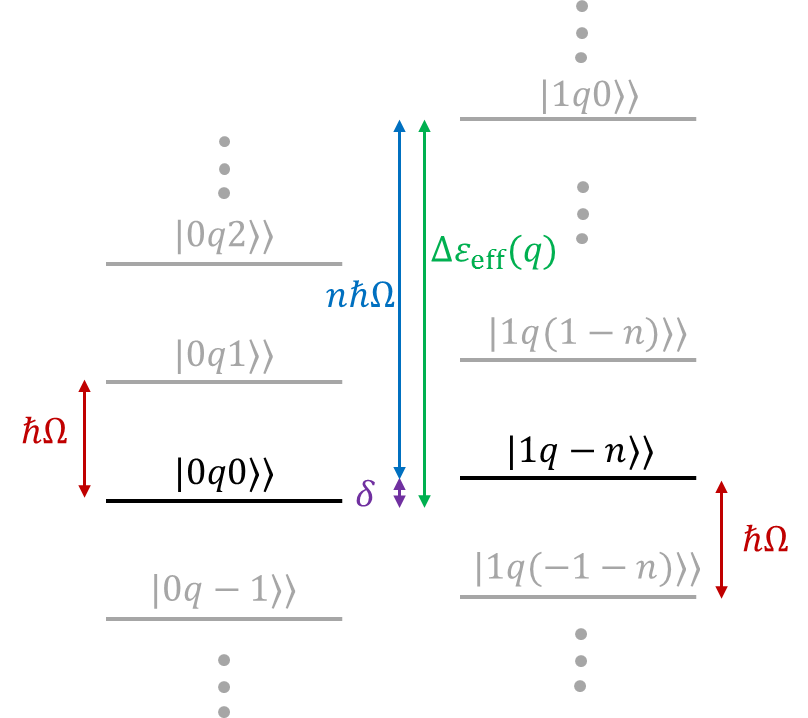}
\caption{(Color online) Resonance condition for an $n$-photon transition between the bands $\alpha=0$ and $\alpha =1$ in the extended Floquet Hilbert space. The energy levels correspond to the unperturbed quasienergies $\varepsilon_\alpha^\text{eff}(q)+m\hbar\Omega$. The states $|\alpha q m\rangle\!\rangle$ are labeled by the band index $\alpha$, the quasimomentum wave number $q$, and the relative ``photon'' number $m$.}
\label{fig:resonance}
\end{figure}
The basis states $|\alpha q m\rra$ correspond to eigenstates $|\alpha q\ra$ of the undriven Hamiltonian. An $m$-photon interband coupling process from state $|\alpha q\ra$ to state $|\alpha' k\ra$ is expected to occur when the resonance condition
\begin{equation}
\Delta E^\text{eff}_{\alpha'\alpha}(q)=\varepsilon^\text{eff}_{\alpha'}(q)-\varepsilon^\text{eff}_\alpha(q) = n\hbar\Omega +\delta
\end{equation}
with integer $n$ and sufficiently small detuning $\delta$ is fulfilled. In the extended Floquet Hilbert space this resonance condition corresponds to a quasidegeneracy of the unperturbed states $|\alpha q m\rra$ and $|\alpha' q (m-n)\rra$ with respect to the diagonal blocks. The energy cost of the transition is compensated by the destruction of $n$ photons. This is illustrated in Fig.\,\ref{fig:resonance} for the case of $\alpha=0$ and $\alpha'=1$, which is relevant for the experiment. Resonant $n$-photon interband excitations are expected when the (effective) coupling matrix element between both states becomes comparable to the detuning~$\delta$. If both states are not coupled directly by a matrix element appearing in the off-diagonal blocks, they might be coupled via an effective matrix element resulting from higher-order processes via energetically distant ``virtual'' intermediate states $|\alpha'' q m''\rra$.

\begin{figure}[t]\hypertarget{fig06ht}{}
\includegraphics[width=0.57\linewidth]{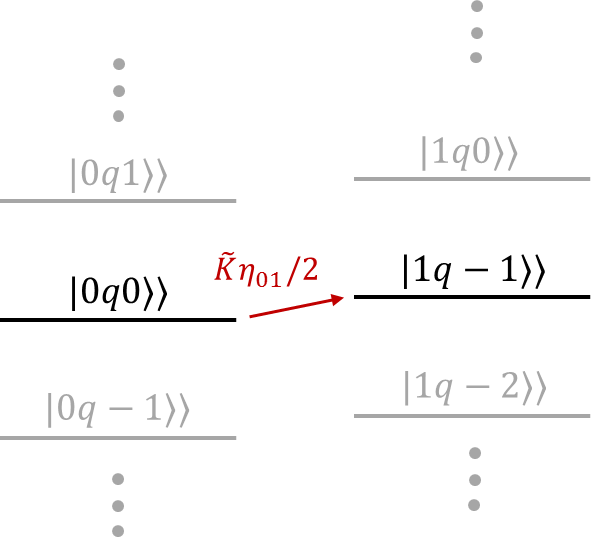}
\caption{(Color online) Single-photon transition.}
\label{fig:single}
\end{figure}

Let us estimate the relevant coupling matrix element for an $n$-photon process from the lowest into the first excited band at quasimomentum $q$, i.e., between the states $|0q\ra$ and $|1q\ra$. For this purpose, we have to evaluate the coupling matrix element between states $|0q0\rra$ and $|1q-n\rra$ in the extended Floquet Hilbert space. For a single-photon process with $n=1$ we find a direct coupling matrix element
\begin{equation}
C_1=\lla 1q-1|\bar{Q}|0q0\rra =  \frac{\eta_{10}}{2}\tilde{K}
\end{equation}
as illustrated in Fig.\,\ref{fig:single}.

\begin{figure}[t]\hypertarget{fig07ht}{}
\includegraphics[width=0.6\linewidth]{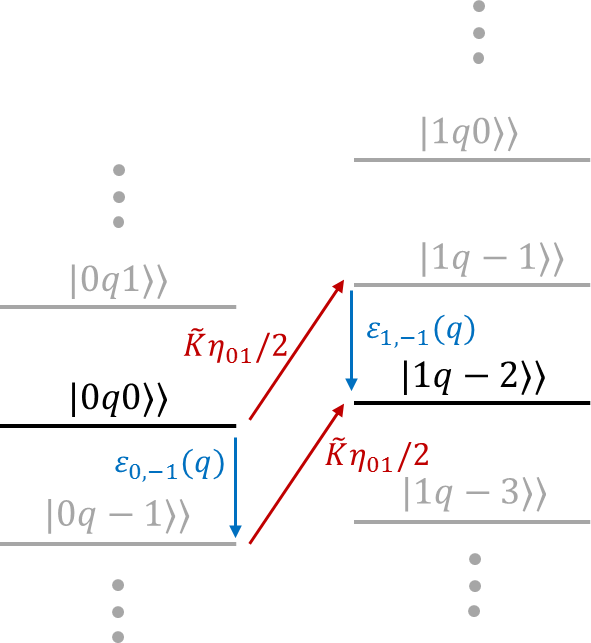}
\caption{(Color online) Two-photon transition.}
\label{fig:two}
\end{figure}

For a two-photon process with $n=2$, we do not find a direct coupling matrix element. However, both states can be coupled via two second-order processes, $|0q0\rra\to|0q-1\rra\to|1q-2\rra$ and $|0q0\rra\to|1q-1\rra\to|1q-2\rra$ as shown in Fig.\,\ref{fig:two}. Using the standard expression of degenerate perturbation theory the effective coupling matrix element is given by
\begin{eqnarray}
C_2&=&\lla 1q-2|\bar{Q}^{(2)}_\text{eff}|0q0\rra
\nonumber\\&=&
  \frac{1}{2}\tilde{K}\eta_{10}\bigg(\frac{\varepsilon_{0,-1}(q)}{\hbar\Omega}
	-\frac{\varepsilon_{1,-1}(q)}{\hbar\Omega}\bigg),
\end{eqnarray}
neglecting $\delta$ in the energy denominator (this level of approximation is equivalent to a high-frequency approximation \cite{Eckardt2015}). The order of magnitude of this term can be estimated by noting that for small arguments the Bessel function behaves like $\mathcal{J}_m(x)\sim x^{|m|}$, such that for $m\ne0$
\begin{equation}
\varepsilon_{\alpha m}(q)\sim \bigg(\frac{\tilde{K}}{\hbar\Omega}\bigg)^{|m|} J_\alpha.
\end{equation}
Including also the momentum dependence of $\varepsilon_{\alpha m}(q)$, which for odd $m$ is just given by a factor of $\sin(aq)$, we find that the matrix element of the two-photon process is of the order of
\begin{equation}
C_2\sim \sin(aq)\bigg(\frac{\tilde{K}}{\hbar\Omega}\bigg)^{2}J_{1,2}.
\end{equation}

\begin{figure}[t]\hypertarget{fig08ht}{}
\includegraphics[width=0.95\linewidth]{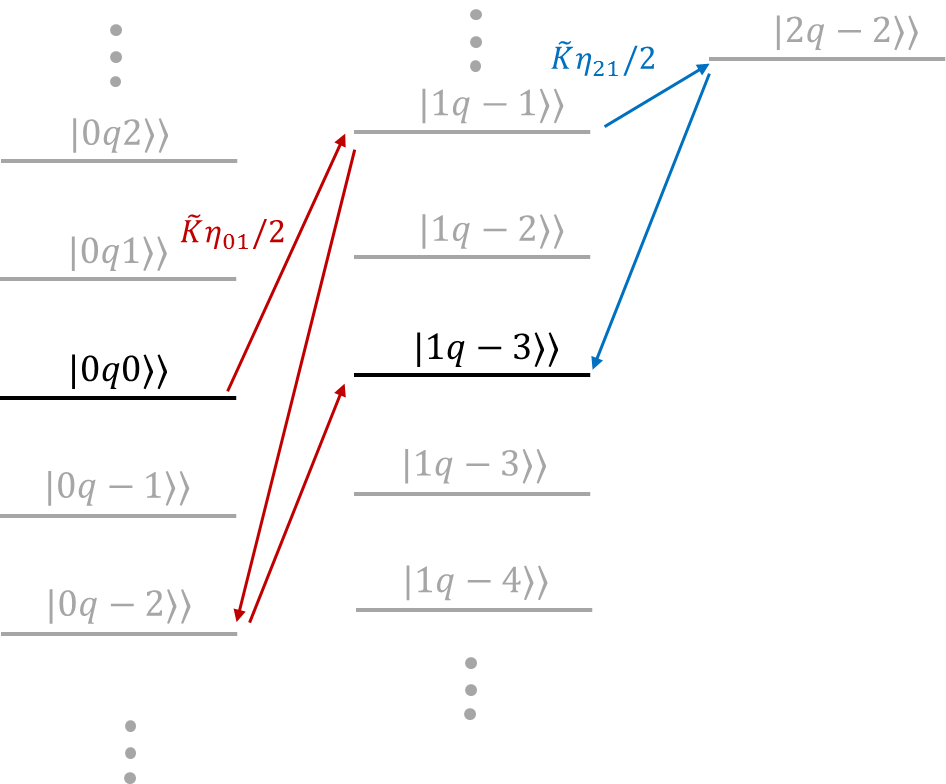}
\caption{(Color online) Three-photon transition.}
\label{fig:three}
\end{figure}

Typical paths contributing to three-photon interband transitions in leading order are depicted in Fig.\,\ref{fig:three}. They also involve intermediate states of higher-lying bands and give rise to effective tunneling matrix elements
\begin{equation}
C_3 = \lla 1 q -3|\bar{Q}^{(3)}_\text{eff}|0q0\rra
\sim  \bigg(\frac{\tilde{K}}{\hbar\Omega}\bigg)^2\tilde{K}.
\end{equation}

Generally, the coupling matrix elements describing an $n$-photon interband transition obey
\begin{equation}
C_n \sim \bigg(\frac{\tilde{K}}{\hbar\Omega}\bigg)^{n-1} \tilde{K}
\quad \text{for odd } n
\end{equation}
and
\begin{equation}
C_n \sim   \sin(aq)\bigg(\frac{\tilde{K}}{\hbar\Omega}\bigg)^{n} J_{\alpha}
\quad \text{for even } n.
\end{equation}
The factor of $\sin(aq)$ results from the fact that for even $n$ always one matrix element $\varepsilon_{\alpha m}(q)$ with odd $m$ always contributes in leading order. [Also higher-order coupling paths for even $n$ contain at least one factor $\varepsilon_{\alpha m}(q)\propto \sin(aq)$ with odd $m$.]
\begin{figure}[t]\hypertarget{fig09ht}{}
\includegraphics[width=1\linewidth]{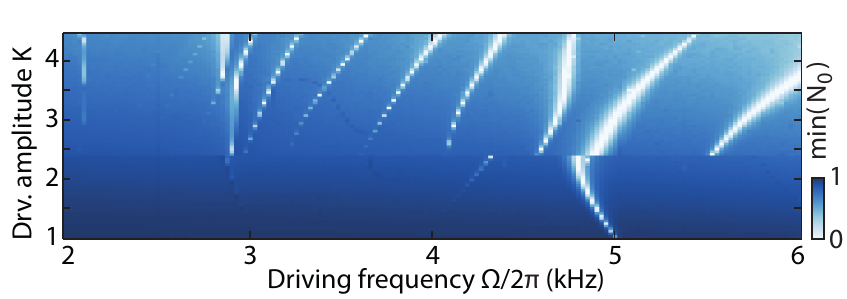}
\caption{(Color online) Minimum occupation of the lowest band during 20 ms of time evolution, versus driving frequency $\Omega$ and
driving strength $K$. At $t=0$ the system is prepared in the Bloch state corresponding to the minimum of the effective
dispersion relation of the lowest band, that is with quasimomentum $q\,{=}\,0$ ($q=\pi/a$) for $K$ lower (larger) than 2.4.}
\label{fig:resonance2}
\end{figure}

We can see that transitions involving an even number of photons are suppressed by an additional factor of $\sin(aq)J_\alpha/(\hbar\Omega)$ with respect to transitions with odd photon numbers. In particular for the experimentally relevant quasimomenta $q=0$ and $q=\pi/a$ resonances with even $n$ are suppressed completely. This is a consequence of the fact that only bands with Wannier functions of different parity are coupled directly by the periodic force on the level of our approximation. The transitions with even photon numbers $n$ are observed experimentally because of the broadening of the quasimomentum distribution due to temperature, interactions, and the finite system extent.

\section{Simulation of the time evolution}
\label{AppendixD}
In order to integrate the time evolution of the shaken lattice, let us start directly from the Hamiltonian in real-space representation (\ref{eq:s12}). As before, we perform a gauge transformation to restore the translational symmetry of the lattice
\begin{equation}
\hat{\mathcal{H}}'(t)= \Uo(t)^\dag\hat{\mathcal{H}}(t)\Uo(t) -\imu\hbar\Uo^\dag(t)\dot{\Uo}(t)
\end{equation}
with
\begin{align}
\Uo(t)= \exp\Big(\hspace{-1mm}-\frac{\imu}{\hbar}\int_0^t\! d t' F_0 x \cos(\Omega t)\Big) = \exp\Big(\imu\Delta q(t) x\Big)
\end{align}
giving
\begin{align}
\hat{\mathcal{H}}'(t)= \frac{1}{2M} [-\imu\hbar\partial_x - A(t)]^2-\frac{V_0}{2}\cos(2\pi x/a),
\end{align}
where $A(t)$ again plays the role of the vector potential. Assuming a system of length $L$ with periodic boundary conditions, we can express the Hamiltonian in terms of momentum eigenstates $|p\ra$ with wave functions
\begin{equation}
\la x|p\ra = \frac{1}{\sqrt{L}} \exp(\imu px).
\end{equation}
For that purpose it is convenient to decompose the momentum wave number as
\begin{equation}
p = q +\beta Q
\end{equation}
with $-\frac{\pi}{a}< q\le\frac{\pi}{a}$, $\beta$ being an integer, and $Q\equiv 2\pi/a$. Introducing the localization energy $E_\text{loc}=\hbar^2\pi^2/(2Ma^2)$, which describes the kinetic-energy cost of localizing a particle on a lattice constant $a$, as the natural unit of energy, we find matrix elements
\begin{equation}
\la q'+\beta'Q|\hat{\mathcal{H}}'(t)|q+\beta Q \ra
	= \delta_{q',q}\,  h_{\beta'\beta}(q,t)E_\text{loc},
\end{equation}
with
\begin{eqnarray}
h_{\beta'\beta}(q,t) &=& \delta_{\beta'\beta} \frac{a^2}{\pi^2}\big[q + \beta Q - A(t)/\hbar\big]^2
\nonumber\\&&+\,\frac{1}{4}\big(\delta_{\beta',\beta+1}+\delta_{\beta',\beta-1}\big)\frac{V_0}{E_\text{loc}}.
\end{eqnarray}
One can see that the wave number $q$ is conserved, so that the dynamics occurs in the space spanned by the integer quantum number $\beta$. By diagonalizing the dimensionless Hamiltonian
$h_{\beta',\beta}(q,t)$ for $A(t)=0$, we obtain the band structure of the undriven lattice. The fact that the diagonal matrix elements $h_{\beta\beta}$ increase like $4\beta^2$, while the off-diagonal terms are constant, $h_{\beta\pm1,\beta}=V_0/(4E_\text{loc})$, shows that Bloch states with energy much larger than the lattice depth resemble plane waves.
\begin{figure}[t]\hypertarget{fig10ht}{}
\includegraphics[width=1\linewidth]{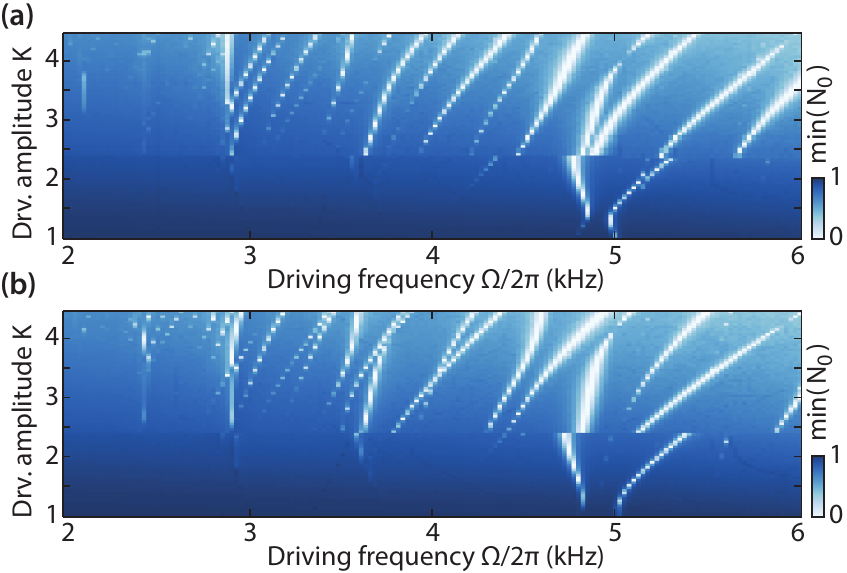}
\caption{(Color online) Same as Fig.\,\ref{fig:resonance2}, but with the initial state shifted away from the minimum by (a) $\Delta q=0.1\pi$ and (b) $\Delta q=0.2\pi$.}
\label{fig:resonance_shift}
\end{figure}
\begin{figure}[t]\hypertarget{fig11ht}{}
\includegraphics[width=1\linewidth]{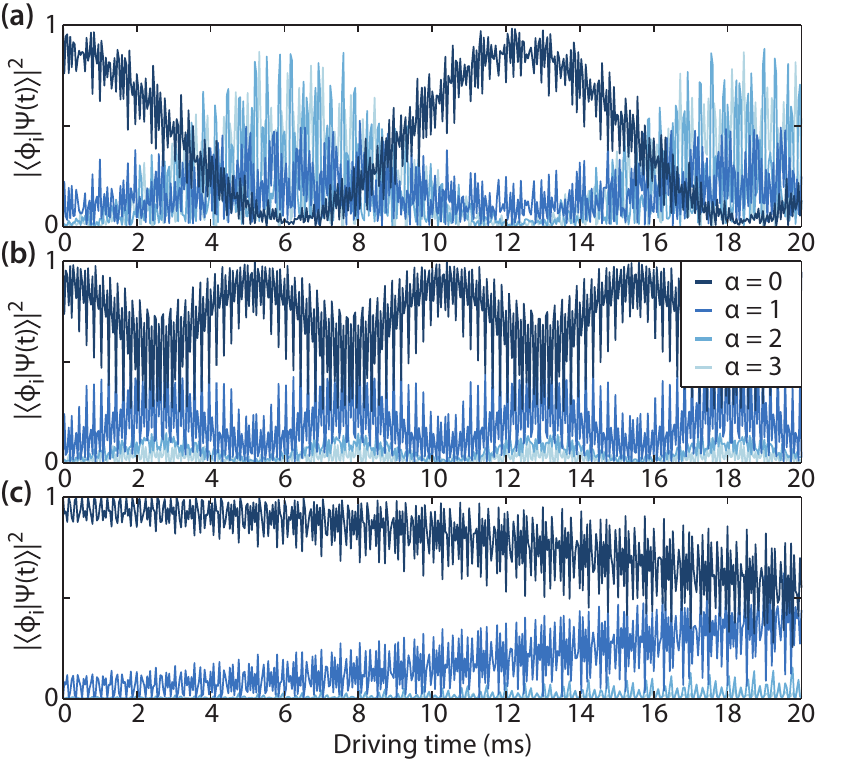}
\caption{(Color online) Occupation of the lowest bands during the time evolution. Plotted is the overlap squared of the time evolved state $\psi(t)$ with eigenstates $\phi_j$ for (a) $K=3.0$ and $\Omega = 2\pi{\times}4.65\,\mathrm{kHz}$, (b) $K=3.0$ and $\Omega = 2\pi{\times}5.00\,\mathrm{kHz}$, and (c) $K=2.5$ and $\Omega = 2\pi{\times}2.90\,\mathrm{kHz}$. The bands for $\alpha =$ 0 to 3 are depicted with increasing brightness (see legend). Bands above $\alpha=3$ exhibit no significant occupation and are omitted from the plots.}
\label{fig:dyn}
\end{figure}

For our simulation we initialize the system in a Bloch state of the lowest band with quasimomentum $q$ and integrate the time-dependent Schr\"odinger equation of the dimensionless time-dependent Hamiltonian $h_{\beta'\beta}(q,t)$ for the given forcing strength $K$ over a time span corresponding to 20\,ms. We take into account 61 plane waves; increasing this number further does not alter the results. We report the minimum occupation of the lowest band encountered during the time evolution. This number provides a measure for the amount of interband excitation expected on the given time scale.

We choose $q$ to be the minimum of the effective dispersion relation of the lowest band, that is $q=0$ ($q=\pi/a$) for $K$ less (greater) than 2.4. With respect to the driving frequency and strength, the minimum occupation of the lowest band is plotted in Fig.\,\ref{fig:resonance2}. While this plot already resembles the experimentally measured data in some respects, it hardly shows resonances corresponding to even photon numbers $n$. This suppression of even resonances is expected from the theory presented in Appendix \ref{AppendixC}.

The even resonances observed in the experiment can be explained by the finite width of the quasimomentum distribution, induced by the finite extent of the trapped system, finite temperature, and interactions. In order to take into account that the quasimomentum distribution possesses a finite width $w$, we simulate the time evolution also starting from initial states that are shifted away from the minimum of the effective dispersion relation by $\Delta q$. As examples, data for $\Delta q=0.1\pi/a$ and $\Delta q= 0.2\pi/a$ are shown in Figs.\,\hyperlink{fig10ht}{\ref{fig:resonance_shift}(a)} and \hyperlink{fig10ht}{\ref{fig:resonance_shift}(b)}, respectively. Here even resonances are clearly visible. The plot shown in Fig.\,\hyperlink{fig03ht}{\ref{fig3}(b)} is a superposition of resonance data obtained for different values of $\Delta q$ (varied in steps of $0.025\pi/a$), with Gaussian weights $\propto\exp(-\Delta q^2/w^2)$.

The width of the momentum distribution was set to $w = 0.1 \pi/a$. While this value cannot be determined experimentally with sufficient accuracy since the time-of-flight pictures shown in Fig.\,\hyperlink{fig01ht}{\ref{fig1}(d)} are taken before the far-field limit is reached \cite{Gerbier:2008bs}, Fig.\,\hyperlink{fig01ht}{\ref{fig1}(d)} still provides an upper bound for $w$, which is of the order of $0.1\pi/a$. An estimate for the lower bound of $w$ is obtained from the Thomas-Fermi radius and is of the order of $0.02\pi/a$ \cite{weinbergSM}. Since thermal and quantum fluctuations will cause further broadening, the value $w=0.1\pi/a$ is a reasonable assumption.

In Fig.\,\ref{fig:resonance2} we can see that the three-photon resonance (near 5\,kHz) from the ground band $\alpha=0$ into the first excited band ($\alpha=1$) is split into two resonances. This is a signature of the fact that the first excited band is coupled resonantly to even higher lying bands, so that an avoided crossing is formed in the quasienergy spectrum. The doubled resonance reflects this avoided crossing and explains the large broadening of the three-photon resonance visible in the experimental data [Fig.\,\hyperlink{fig03ht}{\ref{fig3}(a)}, main text]. In order to identify the bands involved in the three-photon transition, let us have a look at the simulated time evolution. In Fig.\,\ref{fig:dyn} we plot the occupations of the lowest bands over time for different parameters $K$ and $\Omega$ for an initial state with $q=\pi/a$. At the driving strength $K=3$ [see Fig.\,\hyperlink{fig11ht}{\ref{fig:dyn}(a,b)}], the left- and right-hand sides of the double resonance are captured roughly by $\Omega=2\pi{\times}4.65\,\mathrm{kHz}$ and $\Omega=2\pi{\times}5.0\,\mathrm{kHz}$, respectively. We can see that the system performs Rabi-type oscillations between the ground band and a hybridized state with strong contributions from the three bands with $\alpha=$ 1, 2 and 3. We can compare these results with a plot of the time evolution near the five-photon resonance ($K=2.5$, $\Omega=2\pi{\times}2.9\,\mathrm{kHz}$) shown in Fig.\,\hyperlink{fig11ht}{\ref{fig:dyn}(c)}, where only the first excited band becomes populated significantly, since for this resonance an avoided crossings occurs only for larger $K$ (Fig.\,\ref{fig:resonance2}).

\end{appendix}
%

\newpage

\setcounter{figure}{0}
\setcounter{section}{0}
\renewcommand{\thesection}{S\arabic{section}}
\renewcommand{\thefigure}{S\arabic{figure}}
\renewcommand{\thetable}{S\arabic{table}}
\renewcommand{\theequation}{S\arabic{equation}}
\thispagestyle{empty}
\onecolumngrid
\begin{center}
  \large{\textbf{{Supplemental Material}}}
\end{center}
\vspace{4mm}
\begingroup
\leftskip6em
\rightskip\leftskip
\noindent
The following supplemental material provides additional information on the experimental setup and preparation procedure as well as the employed time-periodic driving scheme and data evaluation.
\par
\endgroup
\vspace{10mm}
\twocolumngrid

\section{Experimental setup}
All experiments presented in this paper start with a Bose-Einstein condensate (BEC) of $^{87}\mathrm{Rb}$ atoms in an elliptical crossed optical dipole trap operating at a wavelength of $\lambda_{\mathrm{DT}}=1064\,\mathrm{nm}$. With minimal beam waists of  $\mathrm{w}_{\mathrm{0,h}}=245\,\mathrm{\mu m}$ horizontally and $\mathrm{w}_{\mathrm{0,v}}=82\,\mathrm{\mu m}$ vertically, final trapping frequencies of $\omega_{\mathrm{h}}=2\pi{\times}19\,\mathrm{Hz}$ and $\omega_{\mathrm{v}}=2\pi{\times}48\,\mathrm{Hz}$ are reached after evaporative cooling to quantum degeneracy. Following the creation of a BEC consisting of $(3\pm1){\times}10^5$ atoms in the crossed dipole trap, the 1D optical lattice intensity is linearly ramped up to its final value in $100\,\mathrm{ms}$.
\noindent{
\begin{table}[b]
    \begin{ruledtabular}
        \begin{tabular}{ @{}lccc@{} }
        $N$                                         & $2{\times}10^5$           & $3{\times}10^5$           & $4{\times}10^5$ \\[1ex]
        $N_{\text{Sites}}$                          & $59$                      & $65$                      & $69$ \\
        $n_{\text{max}}$                            & $6.3{\times}10^3$         & $8.7{\times}10^3$         & $1.1{\times}10^4$ \\
        $U_{\text{w}}$      & $1.33{\times}10^{-4}E_{\text{rec}}$     & $1.13{\times}10^{-4}E_{\text{rec}}$     & $1.01{\times}10^{-4}E_{\text{rec}}$ \\
    \end{tabular}
    \end{ruledtabular}
    \caption{\label{tabParameters} System parameters of the one-dimensional lattice with number of atoms $N$, number of occupied lattice sites $N_{\text{Sites}}$, maximum site occupation $n_{\text{max}}$ and occupation weighted on-site interaction parameter $U_{\text{w}}$.}
\end{table}}
With a wavelength of at $\lambda_L=830\,\mathrm{nm}$ the recoil energy of the lattice amounts to $1E_{\mathrm{rec}}=\hbar^2k_L^2/(2M)=h{\times}3.33\,\mathrm{kHz}$, where $M=1.44{\times}10^{-25}\,\mathrm{kg}$ denotes the atomic mass of $^{87}\mathrm{Rb}$. As the two running-wave laser beams of the 1D lattice enclose an angle of $\vartheta=(117.1\pm0.2)^{\circ}$ the resulting lattice constant is given by $a=\lambda_L/[2\sin(\vartheta/2)]= (486.5\pm 0.5)\,\mathrm{nm}$.
For $^{87}\mathrm{Rb}$ atoms in the $|F{=}1,m_F{=}-1\rangle$ hyperfine ground state with an inter-particle scattering length of $a_s=(100.4\pm0.1)a_0$ the Thomas-Fermi profile \cite{Pedri:2001jn} for the given particle numbers and trapping frequencies yields a number of 59-69 occupied lattice sites in pancake-like shapes with increasing particle numbers at the center of the trap (see Tab.\,\ref{tabParameters}). Hereby, the occupation-weighted on-site interaction $U_{\text{w}}=\frac{1}{N}\sum_i^NU_in_i$ ranges from $1.33{\times}10^{-4}E_{\text{rec}}$ to $1.01{\times}10^{-4}E_{\text{rec}}$. Similar experimental conditions are used in Ref.\,\cite{Struck:2012gc}.

\section{Triangular lattice}
The triangular lattice \cite{Becker:2010de,Struck:2011} is comprised of the two laser beams that are also used for the creation of the 1D lattice (denoted beam 1 and 2 in the following) with an additional running-wave laser beam (denoted beam 3) along the $y$ axis. Ideally, all three beams in the $xy$-plane should be aligned at $120^{\circ}$ with respect to each other, yielding wave vectors of $\mathbf{k}_{1/2}= k_L/2\left( \mp \sqrt{3} ,-1,0\right)$ and $\mathbf{k}_3= k_L\left( 0,1,0 \vphantom{\sqrt{3}} \right)$. With each beam being linearly polarized perpendicular to the lattice plane, this setup results in a triangular lattice with a lattice spacing of $a_{\triangleright}=2\lambda_L/3=553.3\,\mathrm{nm}$.
\noindent{
\begin{table}[b]
    \begin{ruledtabular}
        \begin{tabular}{ @{}lll@{} }
        Wave vectors & Angles $\vartheta_{ij}$ \hspace{2mm} & Lattice constants \\[1ex]
        $\measuredangle(\mathbf{k}_1,\mathbf{k}_2)$ & $117.1^{\circ}\pm0.2^{\circ}$ & $\phantomas{|\mathbf{a}_1-\mathbf{a}_2|}{|\mathbf{a}_2|}= 545.9\pm0.8\,\mathrm{nm}$ \\
        $\measuredangle(\mathbf{k}_1,\mathbf{k}_3)$ & $120.4^{\circ}\pm0.4^{\circ}$ & $\phantomas{|\mathbf{a}_1-\mathbf{a}_2|}{|\mathbf{a}_1|}= 560.6\pm1.0\,\mathrm{nm}$ \\
        $\measuredangle(\mathbf{k}_2,\mathbf{k}_3)$ & $122.4^{\circ}\pm0.4^{\circ}$ & $|\mathbf{a}_1-\mathbf{a}_2|= 554.3\pm1.0\,\mathrm{nm}$ \\
    \end{tabular}
    \end{ruledtabular}
    \caption{\label{tabAngles}Measured angles between the lattice vectors $\mathbf{k}_i$ and the corresponding three lattice constants  of the slightly distorted triangular lattice setup [compare Fig.\,\ref{fig4}\textcolor{darkblue}{(a)} in the main text].}
\end{table}}

The resulting scalar lattice potential reads
\begin{align}
    V_{\triangleright}(\mathbf{r}) = V_0\left[ 6+4\sum_{i=1}^3 \cos(\mathbf{b}_i\mathbf{r}-\Delta\phi_{jk})\right],\label{eqnTri}
\end{align}
where $\Delta\phi_{jk}$ are the phase differences between the laser beams with wave vectors $\mathbf{k}_j$ and $\mathbf{k}_k$. The $\mathbf{b}_i$ denote three reciprocal lattice vectors that are given by the circular permutation
\begin{align}
    \mathbf{b}_i = \varepsilon_{ijk}\left( \mathbf{k}_j - \mathbf{k}_k \right)\label{eqnBi}
\end{align}
where $\varepsilon_{ijk}$ is the Levi-Civita symbol, such that $\mathbf{b}_1=\phantomas{\frac{b}{2}}{b} \left(1,\vphantom{\sqrt{3}}0, 0\right)$ and $\mathbf{b}_{2/3}=b/2\left(-1,\mp\sqrt{3},0\right)$ with $b=\sqrt{3}k_L$. Note that the Einstein summation convention is \emph{not} applied in Eq.\,(\ref{eqnBi}). Bravais lattice vectors $\mathbf{a}_i$ in real space can simply be obtained from pairs of reciprocal lattice vectors with the relation $\mathbf{a}_i\cdot\mathbf{b}_j=2\pi\delta_{ij}$ such that, e.g., $\mathbf{a}_1=a_{\triangleright}\left( 0,-1,0 \right)$ and $\mathbf{a}_2=a_{\triangleright}/2\left( \sqrt{3},-1,0 \right)$.

However, as mentioned above, the alignment of the lattice beams is slightly distorted due to a limited optical access in the experimental setup. The experimentally realized angles between the three laser beams can be determined from the position of quasimomentum peaks in time-of-flight images that are generated by a Kapitza-Dirac diffraction and are given in Tab.\,\ref{tabAngles} together with the corresponding lattice constants. For both the 1D lattice as well as the triangular lattice, all calculations of band gap energies shown in the main text are performed with the exact measured lattice beam angles.

Preparation of the atomic ensemble in the triangular lattice follows the same procedure as for the 1D lattice. Since the ensemble is only weakly confined along the $z$-axis (perpendicular to the lattice plane) the atoms form an array of approximately $2100-2600$ elongated cigar-shaped tubes. Similar experimental conditions are used in Ref.\,\cite{Struck:2013ar}. On average the tubes are occupied with 70-95 atoms while the external harmonic confinement results in an occupation of 175-235 in the center of the trap which fades towards the edges of the ensemble.

\section{Sinusoidal Periodic driving}
In the 1D optical lattice, periodic inertial forcing is induced by a frequency modulation $\nu(t) = \nu+\delta\nu(t)$ with $\delta \nu (t)=\nu_0 \sin(\Omega t)$ of one of the two running-wave laser beams. The frequency modulation translates to a phase change of
\begin{align}
    \phi(t)=-2\pi\int_{-\infty}^{\hspace{0.5mm}t}\hspace{-1mm}\mathrm{d}t'\delta\nu(t')
\end{align}
and, thus, a global shift of the rigid lattice potential according to $V_{\mathrm{lat}}\big(\phi(t),\mathbf{r}\big)=V_{\mathrm{lat}}\big(0,\mathbf{r}-\mathbf{R}(t)\big)$ with $\mathbf{R}(t)\,{=}\,\mathbf{\hat{e}}_x\phi(t)/b$ denoting the trajectory of a lattice well. Since $\mathbf{F}(t) = -M\ddot{\mathbf{R}}(t)$ this leads to a periodic forcing of
\begin{align}
    \mathbf{F}(t)=-F_0\cos\left(\Omega t\right)\mathbf{\hat{e}}_x,\hspace{3mm}\text{with}\hspace{2mm}F_0=M\Omega\nu_0 a.\label{eqnForce}
\end{align}
This forcing scheme can be easily extended to the triangular lattice. Here, two of the three lattice laser beams (with wave vectors $\mathbf{k}_{1/2}$) are modulated according to
\begin{align}
    \delta\nu_{1/2}(t)=\pm\nu_x\sin\left(\Omega t\right)+\nu_y\cos\left(\Omega t\right).
\end{align}
The resulting \emph{elliptical} forcing in the triangular lattice is given by
\begin{align}
    \mathbf{F}_\triangleright(t)=-F_x\cos\left(\Omega t\right)\mathbf{\hat{e}}_x-F_y\sin\left(\Omega t\right)\mathbf{\hat{e}}_y\label{eqEllipse}
\end{align}
with independent forcing amplitudes along the $x$ and $y$-axis $F_x=\sqrt{3}M\Omega\nu_xa$ and $F_y=M\Omega\nu_xa$ respectively. Hereby, the prefactor $\sqrt{3}$ in the expression for $F_x$ stems from the different projection of the forcing onto the diagonal lattice directions. Isotropic forcing along all three lattice bonds is, hence, reached if the frequency modulation amplitudes satisfy the condition $\nu_y=\sqrt{3}\nu_x$ as illustrated in Fig.\,\hyperlink{fig04ht}{\ref{fig4}(b)}  in the main text.

\section{Experimental observation of tunneling renormalization}
\begin{figure}[t]\hypertarget{figS1ht}{}
\centering
\includegraphics{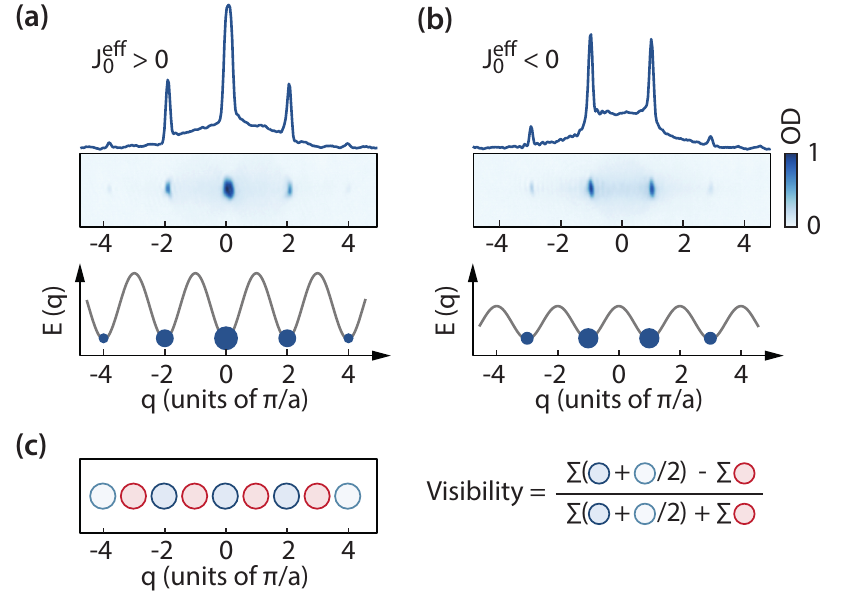}
\caption{(Color online) Quasimomentum distribution of driven 1D lattices. (a) Dispersion relation of a lattice with positive tunneling matrix element and corresponding measured momentum distribution after time-of-flight for a superfluid state. (b) Effective dispersion and measured momentum distribution for negative effective tunneling. The inversion of the dispersion relation is clearly visible in the absorption image by the absence of a zero-momentum component as atoms occupy the edges of the first Brillouin zone. (c) Definition of the momentum \emph{visibility} mask for time-of-flight images of the atomic ensemble in the driven one-dimensional lattice for the characterization of the degree of coherence as well as the effective dispersion relation in the driven system.}
\label{fig:S1}
\end{figure}

The renormalization of effective tunneling matrix elements is essential for the interpretation of the obtained multiphoton spectra. As the presented considerations hold for all energy bands, we omit the band index $\alpha$ in the following. The periodic forcing onto the lattice vectors results in a renormalization of the bare tunneling matrix elements $J_{ij}$ between adjacent lattice sites $i$ and $j$ according to
\begin{align}
    J^{\mathrm{eff}}_{ij}= J_{ij}\frac{1}{T}\int_0^T\mathrm{d}t\exp\left[\imu W_{ij}(t)/\hbar\right],\label{eqnRenorm1}
\end{align}
where $\imu$ denotes the imaginary unit and the projection of the forcing onto the lattice bonds $\mathbf{a}_{ij}$ is given by
\begin{align}
    W_{ij}(t)= -\int_{-\infty}^{\hspace{0.5mm}t}\hspace{-1mm}\mathrm{d}t'\mathbf{F}(t')\cdot\mathbf{a}_{ij}.\label{eqnRenorm2}
\end{align}
Since the tunneling renormalization is chosen to be real-valued and isotropic for all experiments presented within this manuscript, we write $J_{ij}\equiv J$ and $J^{\mathrm{eff}}_{ij}\equiv J^{\mathrm{eff}}$ in the following. For sinusoidal forcing this yields the discussed renormalization of $J^{\mathrm{eff}}= J\mathcal{J}_0\left(K\right)$. For a given lattice potential the dimensionless driving amplitude $K$ depends only on the frequency modulation amplitude. In the case of the 1D optical lattice with periodic forcing along the lattice axis according to Eq.\,(\ref{eqnForce}) all tunneling matrix elements renormalize with
\begin{align}
    K=\frac{aF_0}{\hbar\Omega}=\frac{Ma^2}{\hbar}\nu_0.
\end{align}

Hence, the frequency modulation applied here has the advantage of the tunneling renormalization $\mathcal{J}_0(K)$ being independent of the chosen modulation frequency $\Omega$ in contrast to inertial forcing schemes that rely on direct phase modulation as, e.g., realized by a piezo actuator attached to the retro-reflecting mirror of a counterpropagating lattice setup. The renormalization of the tunneling matrix elements with a zeroth-order Bessel function of the first kind implies that the effective tunneling becomes negative for sufficiently large forcing parameters and, hence, the band structure is inverted. In Fig.\,\ref{fig:S1} both cases of positive- and negative effective tunneling matrix elements are presented [compare Fig.\,\hyperlink{fig01ht}{\ref{fig1}(d)}  in the main text]. Here, the time-of-flight (TOF) images clearly show a coherent occupation of the minima of the effective band structure. For positive tunneling [Fig.\,\hyperlink{figS1ht}{\ref{fig:S1}(a)}] the atomic ensemble exhibits a pronounced quasimomentum component at $q=0$ with integer multiples at $q=\pm2\pi/a$ in accordance with the corresponding Wannier envelope of the momentum distribution. In contrast, the inversion of the effective band structure results in the atomic ensemble residing at the edges of the first Brillouin zone at $q=\pm\pi/a$ for the case of negative effective tunneling without having a quasimomentum component of zero as depicted in Fig.\,\hyperlink{figS1ht}{\ref{fig:S1}(b)}. It is important to note that the total driving time can be experimentally fine-adjusted such that the forcing function $\mathbf{F}(t)$ completes a full period in good approximation before all trapping potentials are shut off for the time-of-flight imaging. At this point in time, the momentum transfer of the forcing onto the atomic ensemble due to the micromotion vanishes such that the oscillating Wannier envelope coincides with the Wannier envelope of a system at rest \cite{Struck:2012gc}.

A convenient quantification of the forcing-dependent behavior of the atomic ensemble is given by the \emph{visibility} as defined in Fig.\,\hyperlink{figS1ht}{\ref{fig:S1}(c)}. The mask applied to the TOF images yields a positive (negative) contribution for an occupation of quasimomentum states corresponding to positive (negative) effective tunneling matrix elements. Hereby the summation scheme over the optical density image pixels ensures equal weighting of positive and negative contributions.

\begin{figure}[t]\hypertarget{figS2ht}{}
\centering
\includegraphics{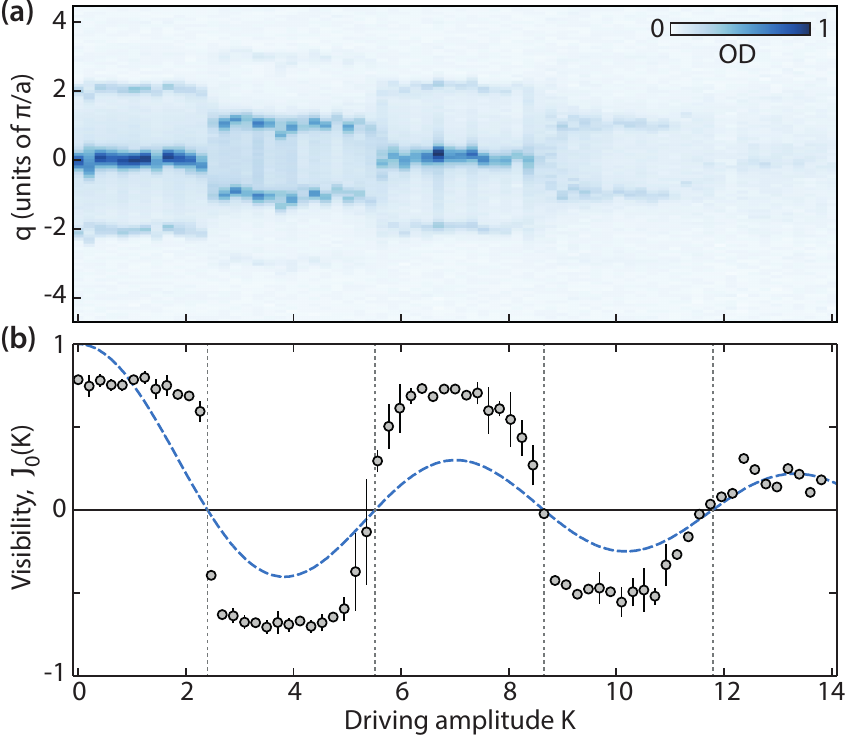}
\caption{(Color online) Periodic driving of the 1D lattice with increasing amplitude. (a) Row-sum of measured TOF-images for increasing forcing parameter $K$. The jump in quasimomentum occupations at the sign-change of the effective tunneling is clearly visible. (b) The zero-crossings of the extracted visibility (data points) coincide remarkably well with the zero-crossings of the renormalization Bessel function (dashed line) that are emphasized by dotted vertical lines. Error bars indicate the standard deviation of the obtained visibility data as each column is averaged over at least two individual TOF-images.}
\label{fig:S2}
\end{figure}

In Fig.\,\ref{fig:S2} the tunneling renormalization is investigated with respect to the forcing strengths. Experiments are performed in a 1D lattice with a lattice depth of $V_0=8.0\,E_{\mathrm{rec}}$. Hereby, the bare tunneling amplitude in the lowest band amounts to $J_0\,{=}\,1.2{\times}10^{-2}\,E_{\mathrm{rec}}$.

As the energy gap between the two lowest bands of the initial dispersion relation is $E_{\text{gap}}\,{=}\,3.6\,E_{\mathrm{rec}}\,{=}\,h{\times}12.1\,\mathrm{kHz}$ the driving frequency is chosen to $\Omega = 2\pi{\times}1.5\,\mathrm{kHz}$. TOF-images have been taken for an increasing final forcing parameter $K$. The column sum of these images is depicted in Fig.\,\hyperlink{figS2ht}{\ref{fig:S2}(a)}, showing a clear and sudden jump between the two distinct cases of positive and negative effective tunneling. This jump is reproduced when extracting the momentum contrast from the images [Fig.\,\hyperlink{figS2ht}{\ref{fig:S2}(b)}]. Zero crossings coincide extremely well with the \emph{ab initio} calculation of the corresponding Bessel function $\mathcal{J}_{\text{0}}(K)$. Remarkably, the measurements still exhibit sharp superfluid momentum peaks even after the third zero-crossing of the Bessel function, indicating that the coherence of the system is retained for very strong driving amplitudes as long as any multiphoton excitations can be avoided by a suitable choice of driving frequency and lattice depth. Exemplified, a forcing parameter of $K=10$ corresponds to a driving amplitude in real-space of approximately five lattice sites.

For the triangular lattice and an elliptical forcing given by Eq.\,(\ref{eqEllipse}) the renormalization can be adjusted independently along the vertical lattice bonds and the two diagonal bonds [denoted $J^{\mathrm{eff}}_{\mathrm{v}}$ and $J^{\mathrm{eff}}_{\mathrm{d}}$, compare Fig.\,\hyperlink{fig04ht}{\ref{fig4}(a)}] by choosing the vertical and horizontal frequency modulation amplitudes $\nu_y$ and $\nu_x$ respectively. The corresponding forcing parameters are
\begin{align}
    K_{\mathrm{v}}=\frac{Ma_{\triangleright}^2}{\hbar}\nu_y\hspace{4mm}\text{and}\hspace{3mm}K_{\mathrm{d}}=\frac{Ma_{\triangleright}^2}{2\hbar}\sqrt{9\nu_x^2+\nu_y^2}.
\end{align}
In Fig.\,\hyperlink{fig04ht}{\ref{fig4}(b)} the renormalization for the diagonal tunneling matrix elements $J^{\mathrm{eff}}_{\mathrm{d}}$ is plotted in dependence of $\nu_x$ and $\nu_y$. The isotropic inversion of all tunneling matrix elements results in a peculiar band structure with two degenerate minima at the vertices of the first Brillouin zone. Note that, due to the reduced total optical density in the time-of-flight images of the driven triangular lattice, driving at the maximum amplitude following the $50\,\mathrm{ms}$ linear ramp is reduced to a duration of $2\,\mathrm{ms}$. Properties of such a driven triangular lattice system are thoroughly discussed in the Refs.\,\cite{Struck:2011} and \cite{Struck:2013ar}.
\begin{figure}[t]\hypertarget{figS3ht}{}
    \centering
        \includegraphics{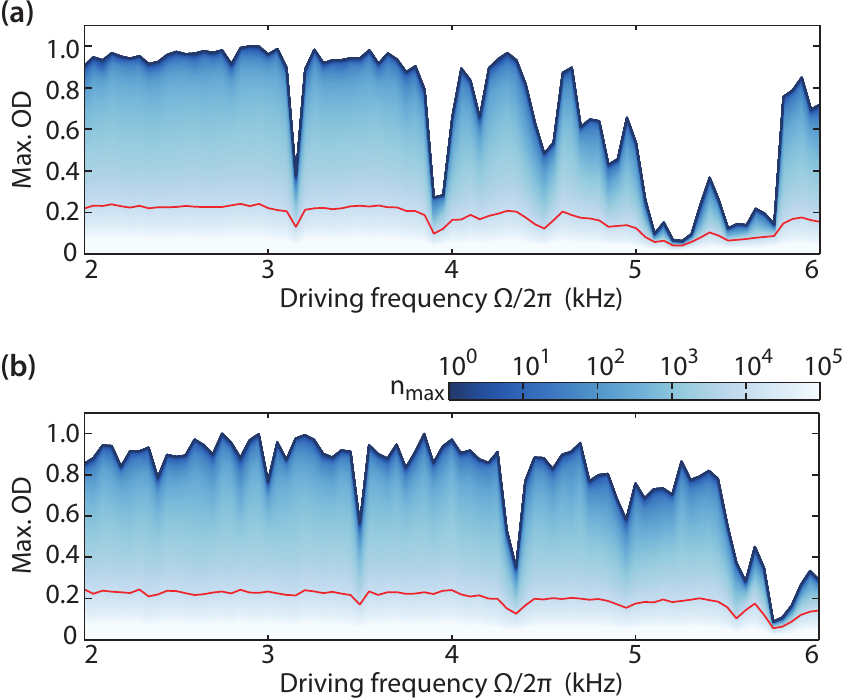}
        \caption{(Color online) Excitation spectra in dependence of the number $n_{\mathrm{max}}$ of maximum-valued pixels included for the calculation of the maximum optical density. Spectra for each value of included pixels are plotted as colored lines corresponding to the color bar. The red solid lines indicate excitation spectra for a number of $n_{\mathrm{max}}=10^4$ included pixels as used for Fig.\,\ref{fig3}\textcolor{darkblue}{(a)} in the main text.}
        \label{fig:S4}
\end{figure}

\section{Data evaluation}
While the visibility defined in the previous section is useful for the characterization of the \emph{effective} tunneling, it cannot easily be applied to the investigation of multiphoton excitations in driven lattices. On the one hand this quantification method has two decisive advantages: To begin with, it yields \emph{per definitionem} a normalized signal due to the summation and normalization over equal surfaces of the absorption image. Furthermore, the obtained visibility value does not only reveal the level of coherence in the system but also gives rise to a sign change of the signal if the effective dispersion relation is inverted in the time-averaged Floquet picture. On the other hand, it can be impractical to account for the periodic motion of the Wannier envelope in momentum space, e.g., by including more momentum components into the mask of Fig.\,\hyperlink{figS1ht}{\ref{fig:S1}(c)} or by fine-adjusting the driving times for varying driving frequencies. Moreover, the quality of the signal is prone to fluctuations. For example the location of the superfluid momentum peaks has to be precisely defined. Finally, the visibility does not yield meaningful results for the case of vanishing effective tunneling (compare Fig.\,\ref{fig2} at $K=2.4$ in the main text).

For these reasons, the maximum optical density has been chosen to quantify the degree of coherence in the system. It is simply obtained by averaging the optical density values for a number $n_{\mathrm{max}}$ of maximum pixels in each TOF image:
\begin{align}
    \mathrm{Max.OD}_{I}\equiv \frac{1}{n_{\mathrm{max}}}\sum_{i=1}^{n_{\mathrm{max}}}p_i\label{eq:maxOD}
\end{align}
with the set of $n_{\mathrm{p}}\gg n_{\mathrm{max}}$ optical density values for each pixel $\mathrm{P}_I=\left\{ p_1,p_2,\ldots,p_{n_{\mathrm{p}}}\right\}$ that are sorted in descending order for each TOF image $I$. Although this method does not yield any information concerning the momentum of the atomic ensemble, it is, thus, an excellent tool for measuring excitation processes for systems that result in an overall loss of coherence.

The number of included pixels that are averaged to obtain the maximum-optical density is arbitrary. A small value of $n_{\mathrm{max}}$ may result in a signal that is very sensitive to even small excitations in the investigated system but can also be prone to detrimental fluctuations such as faulty pixels of the CCD. In addition, the total value of the maximum optical density can also be strongly influenced by the renormalization of tunneling in the effective time-averaged Floquet picture for small $n_{\mathrm{max}}$. While for the excitation spectra shown in Fig.\,\hyperlink{fig02ht}{\ref{fig2}(b)} a value of $n_{\mathrm{max}}=3$ was chosen, the spectra can be easily compared by normalizing the four data sets. For the quantitative comparison of spectra obtained for different driving amplitudes and, hence, different effective tunneling amplitudes as in Fig.\,\ref{fig3}, however, normalization of each single spectrum is not a valid approach. For this reason, a large number of $n_{\mathrm{max}}=10^4$ pixels has been included in the calculation of the maximum optical density in Fig.\,\hyperlink{fig03ht}{\ref{fig3}(a)}. Here, the obtained data yields comparable results even for the zero-crossing of the effective tunneling at $K\approx2.4$.

Despite the stark difference between the number of included pixels in the spectra of Figs.\,\ref{fig2} and \ref{fig3}, their qualitative behavior remains unchanged as shown in Fig.\,\ref{fig:S4}. Here, maximum optical density spectra obtained at a driving amplitude of $K=2.4$ and lattice depths of (a) $V_0=11\,E_{\mathrm{rec}}$ and (b) $V_0=13\,E_{\mathrm{rec}}$ [compare Fig.\,\hyperlink{fig03ht}{\ref{fig3}(b)} in the main text] are plotted over a wide range of values for $n_{\mathrm{max}}$. While, according to Eq.\,(\ref{eq:maxOD}), the absolute values of the maximum optical density necessarily decrease with increasing $n_{\mathrm{max}}$ it is evident that all important features in the spectra remain present without further data processing. This is the case even though for the total number of $n_{\mathrm{p}}\approx 8.5{\times}10^4$ analyzed pixels in each TOF image a value of $n_{\mathrm{max}}=10^4$ pixels corresponds to the inclusion almost $12\%$ of the absorption image in the calculation of the maximum optical density.

\end{document}